\documentclass[final,1p,times,authoryear]{elsarticle}      
\usepackage{fancyhdr}
\usepackage{lineno}

\usepackage[english]{babel}
\usepackage[T1]{fontenc}
\usepackage[utf8]{inputenc}
\usepackage{babel}
\usepackage{amsmath,amssymb,exscale,amsthm}
\usepackage{graphicx}
\usepackage{float}
\usepackage{subcaption}
\usepackage{enumerate}
\usepackage{color}
\usepackage{array}
\usepackage{tabularx}
\usepackage{empheq}
\usepackage{upgreek}
\usepackage{here}
\usepackage{afterpage}
\usepackage{mathabx}
\usepackage{algorithm}
\usepackage{algpseudocode}
\usepackage{slashed}
\usepackage{enumitem}
\usepackage{graphicx}
\usepackage{soul}
\usepackage[labelfont=bf,font=normalsize]{caption}
\usepackage[hidelinks]{hyperref}
\usepackage{natbib}

\DeclareMathAlphabet{\mathbbmsl}{U}{bbm}{m}{sl}
\DeclareRobustCommand{\mathscr}[1]{%
  \text{\usefont{U}{rsfso}{m}{n}#1}%
}

\theoremstyle{definition}

\theoremstyle{plain}

\newtheoremstyle{remarkstyle}{\topsep}
                             {\topsep}
			     {}            
			     {}            
			     {\scshape}    
			     {.}           
			     { }           
			     {\thmname{#1}\thmnumber{ #2}\thmnote{ #3}}
\theoremstyle{remarkstyle}

\usepackage{tikz}
\usepackage{listofitems} 
\usetikzlibrary{arrows.meta} 
\usetikzlibrary{shapes}
\usetikzlibrary{angles,quotes}
\usetikzlibrary{arrows}
\usetikzlibrary{decorations.pathreplacing}
\usepackage[outline]{contour} 
\contourlength{1.4pt}

\tikzset{>=latex} 
\usepackage{xcolor}
\colorlet{myred}{red!80!black}
\colorlet{myblue}{blue!80!black}
\colorlet{mygreen}{green!60!black}
\colorlet{myorange}{orange!70!red!60!black}
\colorlet{mydarkred}{red!30!black}
\colorlet{mydarkblue}{blue!40!black}
\colorlet{mydarkgreen}{green!30!black}
\tikzstyle{node}=[thick,circle,draw=myblue,minimum size=22,inner sep=0.5,outer sep=0.6]
\tikzstyle{node in}=[node,green!20!black,draw=mygreen!30!black,fill=mygreen!25]
\tikzstyle{node hidden}=[node,blue!20!black,draw=myblue!30!black,fill=myblue!20]
\tikzstyle{node convol}=[node,orange!20!black,draw=myorange!30!black,fill=myorange!20]
\tikzstyle{node out}=[node,red!20!black,draw=myred!30!black,fill=myred!20]
\tikzstyle{connect}=[thick,mydarkblue] 
\tikzstyle{connect arrow}=[-{Latex[length=4,width=3.5]},thick,mydarkblue,shorten <=0.5,shorten >=1]
\tikzset{ 
  node 1/.style={node in},
  node 2/.style={node hidden},
  node 3/.style={node out},
}

\numberwithin{equation}{section}
\allowdisplaybreaks

\begin{document}

\begin{frontmatter}

\title{Stochastic Snowball States in Planetary Climate}

\author[1]{Duc Thach Son \textsc{Vu}}

\ead{ducthachson.vu@yale.edu}

\author[2,3]{John S. \textsc{Wettlaufer}\corref{cor}}

\ead{john.wettlaufer@yale.edu, jw@fysik.su.se}

\cortext[cor]{Corresponding author: \texttt{john.wettlaufer@yale.edu, jw@fysik.su.se}}

\address[1]{Department of Earth \& Planetary Sciences, Yale University, New Haven, CT 06511, USA}
\address[2]{Departments of Earth \& Planetary Sciences, Mathematics and Physics,
Yale University, New Haven, CT 06511, USA}
\address[3]{Nordita, KTH Royal Institute of Technology and Stockholm University, SE-106 91, Stockholm, Sweden}

\begin{abstract}

Motivated by the recent finding of \citet{JiAbbot25} that at moderate eccentricity (0.2--0.3) the planetary Snowball bistability vanishes, we have introduced stochasticity of the orbital eccentricity into a simplified version of the energy balance model of \citet{EW09} to study a general Earth-like planet. The stochastic eccentricity is approximated by two dominant time scales, annual and daily, and is introduced through an analysis of the Kepler equation. We find that, apart from mean eccentricities less than 0.1, the stochastic dynamics are quantitatively and qualitatively different than the deterministic dynamics.  For a fixed mean eccentricity, fluctuations in eccentricity amplify the asymmetry and intermittency in radiative flux forcing, thereby increasing the time it takes to reach a stable seasonal cycle.  As the mean eccentricity increases above 0.1, the stochastic transition from the Snowball to non-Snowball state occurs over a range of mean stellar flux $\Delta S^*$ rather that at a particular stellar flux $S^*$ for the deterministic case. Moreover, as the mean eccentricity increases from $0.1$ and the variance of the annual eccentricity fluctuations increases, $\Delta S^*$ increases and the probability of the system reaching a steady state and experiencing a Snowball to non-Snowball bifurcation decreases.
This implies that there may be many high eccentricity planetary systems in a perpetually transient state. 

\end{abstract}






\end{frontmatter}

\section{Introduction}\label{se:intro}

The climate on any Earth-like planet is controlled by solar energy input, longwave radiative output and the internal redistribution of energy by the atmosphere and the ocean \citep{Pierrehumbert2010}. On Earth, of the 5 PW excess energy input at low latitudes about two thirds is transferred poleward by the atmosphere and the remainder by the ocean.  \citet{Budyko1969} and \citet{Sellers1969} constructed energy balance models that ask the general question of how an ice-covered and ice-free Earth may be stable states that could coexist under the same climate forcing. The essential physics underlying such a situation is the ice-albedo feedback, with the high albedo low temperature ice state persisting due to the reflection of incident solar radiation, and the low albedo ocean persisting due to absorption of incident solar radiation.  The intermediate, partially ice-covered, state is unstable to areal fluctuations beyond some threshold and a saddle-node bifurcation drives the system to either of the stable states, defining a hysteresis loop that has long been fundamental to planetary climate science \citep[e.g.,][]{NorthCahalanCoakley1981}. \cite{Budyko1969} and \cite{Sellers1969} predicted that a state of global glaciation could be reached on Earth.  Such a ``Snowball Earth'' prediction eventually received confirmation in the geological record \cite[See][and Refs therein]{Kirschvink1992LateProterozoic, HoffmanEtAl1998SnowballEarth, Pierrehumbert2011ClimateNeoproterozoic}. Here we examine the role of stochastic orbital eccentricity in Snowball states in a simplified model of an Earth-like planet.

\subsection{Motivation from Earth}

Earth's orbit and tilt underlie the strong seasonal cycle of climatological data; climate variability is typically viewed in terms of fluctuations in the seasonal cycle induced by higher frequency processes, which represent the competition between the orbitally enforced monthly stability and the fluctuations driven by weather.  Clearly, the weather is high frequency in comparison to the monthly time scale, the monthly time scale is high frequency relative to the seasonal time scale,...,the millennial time scale is high frequency relative to the glacial cycles, and so forth.  The default approach is to compare time scales qualitatively, and by standard power spectral analysis of time series, typically linking them with heuristic arguments and/or complex climate models.  
However, stochastic methodologies can demonstrate analytically what cycles are important on which timescales, {\em and} reveal the nature of the coupling between the core phenomena operating on these time scales.  

Motivated by Munk's analysis \citep{Munk1960} of observations of the spectral properties of pressure fields using methods of signal processing \citet{Mitchell1966} posited an autonomous stochastic differential equation description of the ocean climate. 
\citet{Hasselmann1976} derived a stochastic description of ocean climate in which the ``noise'' is associated with the chaotic weather fluctuations as described by \citet{Lorenz1963}. In direct analogy with Brownian motion, he asked the general question of how high-frequency fluctuations in the weather impact low-frequency climate states, which is often referred to as the ``Hasselmann Program'' \citep{LucariniChekroun2023}.

In their spectral analysis of instantaneous eccentricity fluctuations of solar system planets from 13000 BC to 17000 AD, \citet{Scafetta19} found significant power in the Earth-Moon system on time scales of 0.6152, 1.2029, 2.671, 3.983, 5.931, 8.10, 11.86, 15.78 and 29.43 years, with both higher and lower frequency contributions at lower amplitude.

\citet{Laskar2011} present long-term numerical solutions for Earth’s orbital eccentricity, which they argue is suitable for paleoclimate applications back to about 50 Myr, beyond which chaotic divergence limits predictability. They highlight the robust eccentricity signatures, especially the 405 kyr cycle and its 2.4 Myr modulation, and importantly they present eccentricity realizations that exhibit stochasticity with some 60\% variation.  

Clearly eccentricity is coupled to the other orbital cycles, the relative roles of which in Earth's glaciations have been debated \citep{huybers2005obliquity, huybers2011,Hobart2023Precession100kyr}. For example, \citet{Barker2025} argued that over the past 900 kyr, orbital forcing alone can explain both the occurrence and duration of all deglacial and interglacial periods on Earth.  Their analysis is consistent with glaciations occurring when eccentricity decreases due to the associated reduction in the amplitude of precession, or by parity of reasoning, deglaciations are facilitated by larger eccentricity and precession amplitude \citep{huybers2011}.  The ongoing debate concerns the relative roles of internal climate dynamics versus external forcing \cite[e.g.,][and refs therein]{Zhang2025}.

In a recent example, \citet{Wu2024} used the CAM3–CLM3 Global Climate Model (GCM)  coupled to an explicit meltpond scheme to examine how variations in Earth’s orbital parameters modulate the CO$_2$ level required to trigger Snowball Earth deglaciation.  They find that the orbital configuration can shift the deglaciation threshold substantially, with the most favorable configuration having high eccentricity, low obliquity, and precession placing an equinox near perihelion, wherein they find that the threshold partial pressure of CO$_2$ is ~40\% lower than under modern orbital conditions.

\subsection{Motivation from Earth-like exoplanets\label{se:exo}}

From binary and triple stars, up to globular clusters with some $10^7$ stars, stars form systems of different multiplicity \citep{KiselevaEggletonEggleton2001}.  Importantly, observations yield a frequency of binary and multiple star systems in the Galactic field of up to $70 \%$, where between 5 and $15 \%$ are at least triple \cite[See][and Refs therein]{KiselevaEggletonEggleton2001}.  A hierarchical triple system consists of a binary system and a third body on a wider orbit, the motion of which can be pictured as the motion of inner binary itself, and the binary consisting of the third body relative to the center of mass of the inner binary.  Complex three-body problems where the third body is a planet--a circumbinary planet--form an example of interest here.

\citet{Doyle2011Kepler16} discovered the first clear transiting circumbinary planet, Kepler-16 b, orbiting an eclipsing binary star. The planet transits both stars and the orbital event timings and geometry can only be explained using a 3-body configuration.  The circumbinary transits do not follow a simple ephemeris because the apparent ``period'' between observed planetary transit-like events is not constant.  This is due to the fact that the stars move significantly during the planet’s orbit; the planet meets each star at different binary phases.  Thus, because of the large secondary-to-primary mass ratio and orbital eccentricity of the binaries, circumbinary planet orbits are significantly non-Keplerian.
The Saturn sized planet is on a nearly circular 229-day average orbit around its two parent stars, which have an eccentric 41-day orbit. However, secular variations occur on a characteristic timescale of roughly 40 years--the planet’s eccentricity can evolve upward to about 0.09 even though its fitted value at discovery epoch is near zero.  Additionally, over an orbital timescale the eccentricity can vary by over a factor of two or three and thus a near-zero value at one epoch is expected physically (see their Figure S3).  They find that the best-fit 3-body solution integrated numerically forward 2 Myr is stable on that time scale.

In their study of the three circumbinary planets, Kepler-16 b, Kepler-34 b and Kepler-35 b, \citet{LeungLee2013AnalyticTheory} analyze the origin of the high and low frequency oscillations of eccentricity.  They find that the slower apsidal beats of these three circumbinaries have periods of 
$P_{\rm aps}\approx 42~{\rm yr},\;91~{\rm yr}$ and $20~{\rm yr}$ respectively.  Overlain upon this apsidal evolution are high frequency fluctuations associated with the guiding-center and the epicycle, or free-eccentricity, dynamics with periods $P_0$ and $P_\kappa$ respectively. These are sub-year across the three circumbinary planets with a range of $P_0, P_\kappa \approx 0.35\text{ to }0.78~{\rm yr}$.  
Therefore, when eccentricity is computed from the instantaneous state (osculating elements), it is natural to observe high frequency eccentricity oscillations on multiple sub-year timescales because of beating between these processes, plus additional forced harmonics, as seen in their Figures 1, 4, 5 and 7, which show high frequency eccentricity variations from a factor of two to eight depending on the circumbinary system (cf. Figure S3 of \cite{Doyle2011Kepler16}).  See the Supplementary Material in \ref{sec:SM} for a detailed description of the material in \citet{LeungLee2013AnalyticTheory} underlying these periods.

\citet{PoppEggl2017CircumbinaryClimate} used a GCM to study an aquaplanet with a fixed CO$_2$ Earth-like atmosphere and orbital variations motivated by Kepler-35 b.  They determine the top-of-atmosphere radiative forcing from the instantaneous circumbinary star–planet distances, which in their co-planar orbital model \citep{GeorgakarakosEggl2015AnalyticOrbitPropagation} has a period of about 22 Earth-days, associated with the center of mass rotation of the two stars, and a period of about 360 Earth-days, associated with eccentricity of the planetary orbit.  
They also note that there are short but substantial eclipse-related insolation reductions about every 11 Earth-days.  Their goal was to assess how variations in insolation affect the mean climate and habitability of the Earth-like circumbinary planet.  They define their planet as uninhabitable when either the surface is completely ice covered, or when the upper atmospheric water vapor concentration exceeds the Moist Greenhouse limit, beyond which water vapor is lost.  They conclude that their approximate orbitally driven insolation produces stronger climate variability, such as spectral peaks and seasonal-like structure, but does not substantially shift their defined mean habitable range for the scenarios they examined.

Recently \citet{JiAbbot25} used a fixed range of orbital eccentricity in the GCM ExoCAM and 
in the energy balance model of \citet{EW09} to assess the bi-stability of the planetary snowball state.  They applied the energy balance model in the equatorial region of a Snowball climate state with an annual cycle in stellar forcing due to an eccentric orbit. 
 The favorable comparison between these two approaches shown in their Fig. 3 motivated them to use the energy balance model to interpret and explain their results. They found that the bi-stability vanishes when the orbital eccentricity is increased to 0.2–0.3, which is within the range found in the NASA exoplanet archive (See their Fig. 1 and the Supplementary Information in \ref{sec:SM}). They noted that, upon accounting for observational uncertainties, 33\% of rocky exoplanets have $1 \sigma$ upper limits on eccentricity that exceed 0.25, whereas at higher eccentricities stabilizing perennial ice becomes unlikely.
Their choice of eccentricity within this range was static in their analysis.    

A different general mechanism of eccentricity variations originates in the discovery of \citet{Poincare1890} of the architecture of chaos in the restricted three-body problem through the homoclinic tangle of orbits, which in general underlies multiple timescales in orbital dynamics.  For example, \citet{Wisdom85} showed the perturbative effects of asteroid dynamics near Jupiter’s 3:1 mean motion resonance in the planar elliptic restricted three-body problem.  By separating orbital, resonant libration, and secular precession time scales he explained the characteristic alternation between low  and high eccentricity behavior in chaotic trajectories, with bursts of eccentricity of a factor of four on a millenial time scale (see his Fig. 2). 

Motivated by these and related studies, we simplify the stochastic nature of orbits, which takes place on multiple time scales due to intrinsic many body processes.  In so doing, we study a general Earth-like planet by introducing a minimal two-timescale stochastic eccentricity into the energy balance model \citet{EW09} to examine the influence of fluctuations in insolation on the bifurcation structure of a planetary snowball state. 

\section{Energy balance model and bifurcation of the deterministic ice-ocean system}

In the model of \citet{EW09} (EW09) the state variable $E$ represents the energy per unit area stored in sea ice as latent heat when the ocean is ice-covered or in the ocean mixed layer as sensible heat when the ocean is ice-free,
\begin{equation}
E=\left\{
\begin{array}{ccl}
     -L_i h_i & E<0 & [\text{Sea ice}]\\
     c_{\text{ml}}H_{\text{ml}}T_{\text{ml}} & E\geq 0 & [\text{Open ocean}] 
\end{array}
\right. ,
\end{equation}
where $L_i$ is the latent heat of fusion for sea ice, $h_i$ is the ice thickness, $c_{\text{ml}}$ is the mixed-layer specific heat capacity, and $H_{\text{ml}}$ is the mixed-layer depth. The ocean mixed-layer temperature $T_{\text{ml}}=\tilde{T}_{{\text{ml}}}-\tilde{T}_{\text{fr}}$ is measured relative to the freezing point $\tilde{T}_{\text{fr}}$, taken to be $0^\circ$, where $\tilde{T}_{\text{ml}}$ is the actual mixed layer temperature. The time evolution of $E$ is governed by a first-order nonautonomous ordinary differential equation
\begin{equation}
  \frac{\mathrm{d} E}{\mathrm{~d} t}=f(t, E),  
\end{equation}
wherein the complete time varying energy balance, $f(t, E)$, involves the ice-albedo dependent difference between the incoming shortwave radiation at the surface and the outgoing longwave radiation, augmented by sensible and latent heat fluxes and an additional amount associated with greenhouse gas forcing. The Stefan-Boltzmann equation for outgoing longwave radiation is linearized in the deviation of the ice surface temperature from the freezing point, and the seasonally varying values of surface fluxes are modulated by an atmospheric model incorporating observations of cloudiness, atmospheric transport from lower latitudes and the meridional temperature gradient.  

\citet{MoonWettlaufer2017StochasticSeaIce} developed a stochastic version of EW09 and examined the roles of additive and multiplicative noise on the bifurcation structure of Arctic sea ice with increasing greenhouse gas forcing.  
They found that the stochasticity fundamentally influences the nature of the deterministic steady-state solutions corresponding to perennial ice, seasonal ice and ice-free states. Their approach motivates our treatment of stochastic orbital eccentricity.  

Simplified variants of EW09 were investigated by \citet{E12} and \citet{Hill16} as
\begin{equation}
    \dfrac{dE}{dt}=F-BT_s,
    \label{eq:simpODE}
\end{equation}
where $B$ is constant and the surface temperature $T_s$ is defined as
\begin{equation}
T_s=\left\{
    \begin{array}{lll}
    \dfrac{E}{c_{\text{ml}} H_{\text{ml}}}& E\geq 0 &[\text{open ocean}]\\
    0                  & E<0,F>0&[\text{sea ice, melting surface}]\\
    \dfrac{F}{B}\bigg(1+\dfrac{k_i}{Bh_i}\bigg)^{-1} &E<0,F<0 &[\text{sea ice, frozen surface}]
\end{array}\right.
\label{eq:simpT}
\end{equation}
\citet{JiAbbot25} used this approach to investigate the transition from the Snowball to the non-Snowball state at moderate orbital eccentricity when the ice thickness $h_i$ vanishes at some time in the seasonal cycle.  This akin to the reversible transition from the perennial Arctic ice cover to the seasonal ice cover analyzed by \citet{EW09} (see their Fig. 3). 

We rewrite Equations \eqref{eq:simpODE} and \eqref{eq:simpT} as
\begin{equation}
\begin{array}{ll}
     \dfrac{dE}{dt} & =\text{Absorbed Stellar flux + Advective Flux - Outgoing Longwave Radiation (OLR)} \\
     & =\dfrac{(1-\alpha_i)S_0}{\pi r^2(t)}+Q_{adv}-[A+BT_s(t)]\\
     & =F(t)-B T_s(t),
\end{array}
\label{eq:system_main}
\end{equation}
where $F(t)=\dfrac{(1-\alpha_i)S_0}{\pi r^2(t)}+Q_{adv}-A$, $\alpha_i$ the top of the atmosphere (TOA) albedo with an ice-covered surface, $S_0$ the stellar flux at 1 AU, $r$ the distance from the planet to the star in astronomical units (AU), and $T_s$ is the surface temperature in Celsius. The orbital distance $r$ depends on eccentricity and can be written as a function of the eccentric anomaly, $\psi$, as $r=a(1-e\cos \psi)$ where $a$ is the semi-major axis (in AU) and $e$ is the eccentricity.  Kepler’s equation, $M=\psi-e\sin \psi$, provides the value of $\psi$ through the time-dependent mean anomaly $M$. The advective flux is scaled to the annual-mean stellar flux as $Q_{adv}=Q^*_{adv}+(S/S_0-1)\Delta Q_{adv}$ where $S=S_0/(a^2\sqrt{1-e^2})$. The parameters used to close the model are defined in Table 1, which are the same as used by \cite{JiAbbot25}, and we also systematically vary the solar flux from the nominal value $S_0$ as described below.

\def\thickhline{\noalign{\hrule height.9pt}}
\newcolumntype{\thickvertical}{!{\vrule width 1pt}}
\begin{table}[H]
\begin{center}
\caption{Values of parameters}\label{tab:parameters}
\begin{tabular}{ \thickvertical l|c|c|l \thickvertical} 
\thickhline
Parameter & Symbol & Value & Unit \\
\thickhline
Stellar flux at 1 AU & $S_0$ & 1360 & Wm$^{-2}$\\
Latent heat of fusion of ice & $L_i$ & 9.5 & Wm$^{-3}$yr\\
Ocean mixed layer heat capacity times depth & $c_{\text{ml}}H_{\text{ml}}$ & 6.3 & Wm$^{-2}$yrK$^{-1}$\\
Ice thermal conductivity & $k_i$ & 2 & Wm$^{-1}$K$^{-1}$\\
OLR constant & A & 173 & Wm$^{-2}$\\
OLR coefficient & B & 1.44 & Wm$^{-2}$K$^{-1}$\\
Equatorial advective heat flux constant & $Q^*_{adv}$ & -22 & Wm$^{-2}$\\
Equatorial advective heat-flux coefficient & $\Delta Q_{adv}$ & -100 & Wm$^{-2}$\\
TOA albedo with ice-covered surface & $\alpha_i$ & 0.55 & --\\
\thickhline
\end{tabular}
\end{center}
\end{table}

\citet{JiAbbot25} investigated the bifurcation from the Snowball to the non-Snowball state by varying $S_0$, among other parameters. A criterion for bifurcation was the vanishing of ice thickness, $h_i=0$, or equivalently $E=0$. As noted above, their simulations assumed a fixed value of eccentricity.  Depending on the time scales involved, this assumption may be valid.  

Here, we adjust $S_0$ with a scale factor $S^*\times S_0$. A bifurcation occurs when the system converges to a stable solution that is a non-Snowball state: the absolute difference in $E$ between the beginning and the end of a given orbital year does not exceed 0.1 Wm$^{-2}$yr. If the maximum value of $E$ in such an orbital year is nonnegative, then a bifurcation occurs. In \ref{se:E_decrease}, we determine the critical values of $S^*$ such that $E$ decreases with time; the deterministic system \eqref{eq:system_main} requires a sufficiently large stellar flux to drive the transition from a Snowball state to a non-Snowball state.  In the next section we examine the implications of the eccentricity being governed by simple Gaussian noise as manifested through Kepler's equation.


\section{Analysis of stochastic eccentricity in the ice-ocean energy balance model}\label{se:noise}

Motivated by the multiple timescales in Earth-like exoplanets and circumbinaries described in \ref{se:exo}, now we investigate the effect of stochastic fluctuations in the eccentricity and its influence on other quantities in equation \ref{eq:system_main} through the well-known Kepler equation. This simplification maintains the framework of the non-autonomous stochastic differential equation \ref{eq:system_main} at the expense of the specificity of a particular circumbinary system, all the while introducing the multiple timescales in eccentricity exhibited by all of the circumbinaries discussed.


\subsection{Kepler's equation and stochastic eccentricity}

\begin{figure}[H]
    \centering
    \includegraphics[width=0.8\linewidth]{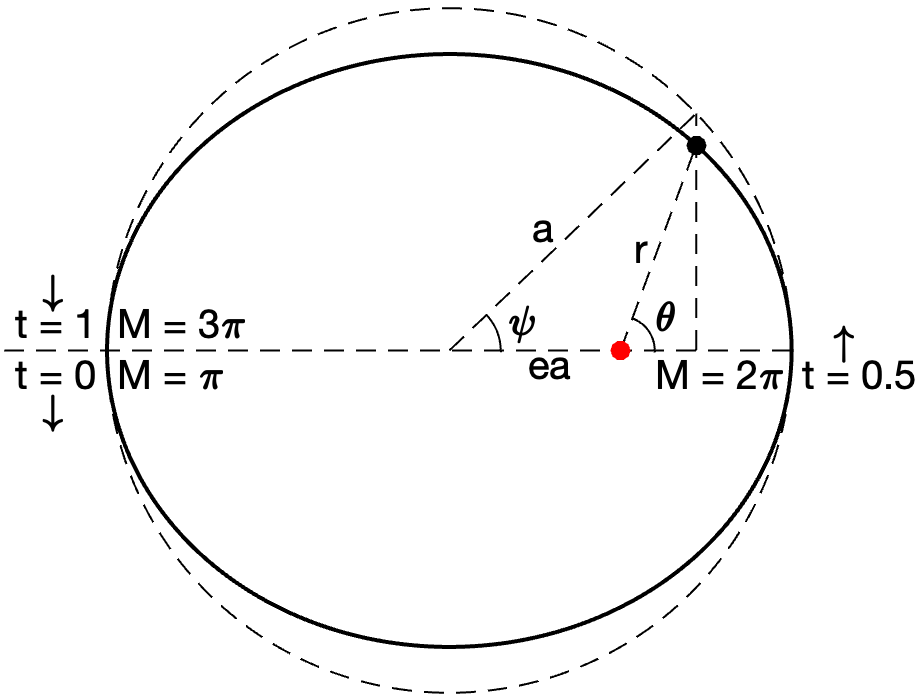}
    \caption{Elliptical orbit determined from Kepler’s equation; $M=\psi-e\sin \psi$, linking the mean anomaly $M$, the eccentric anomaly $\psi$, and the eccentricity $e$. The red point represents the central star, and the black point at $r$ represents the planet, with semi-major axis $a$.}
    \label{fig:kepler_orbit}
\end{figure}

Kepler’s equation $M=\psi-e\sin \psi$ is central to orbital dynamics, linking the mean anomaly $M$ and the eccentric anomaly $\psi$, thereby governing the temporal evolution of bodies moving along elliptical trajectories, as illustrated in Figure \ref{fig:kepler_orbit}.  We start every orbit ($t=0,1,2,...$) in winter at the farthest point in the counterclockwise orbit of trigonometric anomaly angles.  For example, $M=\psi=\pi$ at the beginning of the first orbit $t=0$; $M=\psi=2\pi$ in mid-orbit $t=0.5$, and $M=\psi=3\pi$ at $t=1$, the beginning of the second orbit. The general formula is $M=\big[\{t\}+0.5\big]2\pi$, where $\{t\}$ is the mantissa of $t$.


The solutions of Kepler’s equation have been the subject of extensive study for centuries, despite the absence of a closed-form expression in terms of elementary functions. However,  numerous analytical approximations, asymptotic expansions, and numerical schemes have been developed to address this fundamental problem. In \ref{se:small_variance}, we analyze the distributions of 
$e, \psi, \cos \psi, r, r^2$ and $1/r^2$ and obtain an approximate Gaussian distribution for $1/r^2$ with sufficiently small variance around a mean eccentricity.

\subsection{Introducing fluctuating eccentricity into the ice-ocean energy balance model}

Using the results in \ref{se:small_variance}, with sufficiently small standard deviation $\sigma$ and Gaussian white noise $\xi(t)$ with a normal distribution $N(0,\sigma_E)$ ($\sigma_E \ll 1$), we set $\dfrac{1}{r^2(t)} \rightarrow \dfrac{1}{r^2(t)}+\xi(t)$ in equation \ref{eq:system_main} giving 
\begin{equation}
\begin{array}{ll}
\dfrac{dE}{dt}& =\dfrac{(1-\alpha_i)S_0}{\pi}\bigg[\dfrac{1}{r^2(t)}+\xi(t)\bigg]+Q_{adv}-A-BT_s\\
& = F(t)-BT_s+\dfrac{(1-\alpha_i)S_0}{\pi}\xi(t)\\
& = f(E,t)+c\xi(t),
\end{array}
\label{eq:system_stoch}
\end{equation}
where $f(E,t)=F(t)-BT_s$ and $c=(1-\alpha_i)S_0/\pi$ is a constant. 
When $\sigma$ and $\sigma_E$ are sufficiently small, then $\xi(t)\ll1$ at every time $t$, and the stochastic solution does not deviate from the deterministic solution of $\dfrac{dE}{dt}=f(E,t)$. However, as the variance $\sigma^2$ increases we expect, and find in our numerical simulations, that the stochastic and deterministic solutions deviate.

\begin{figure}[H]
    \centering
    \begin{subfigure}{0.49\textwidth}
        \centering
        \includegraphics[width=\textwidth]{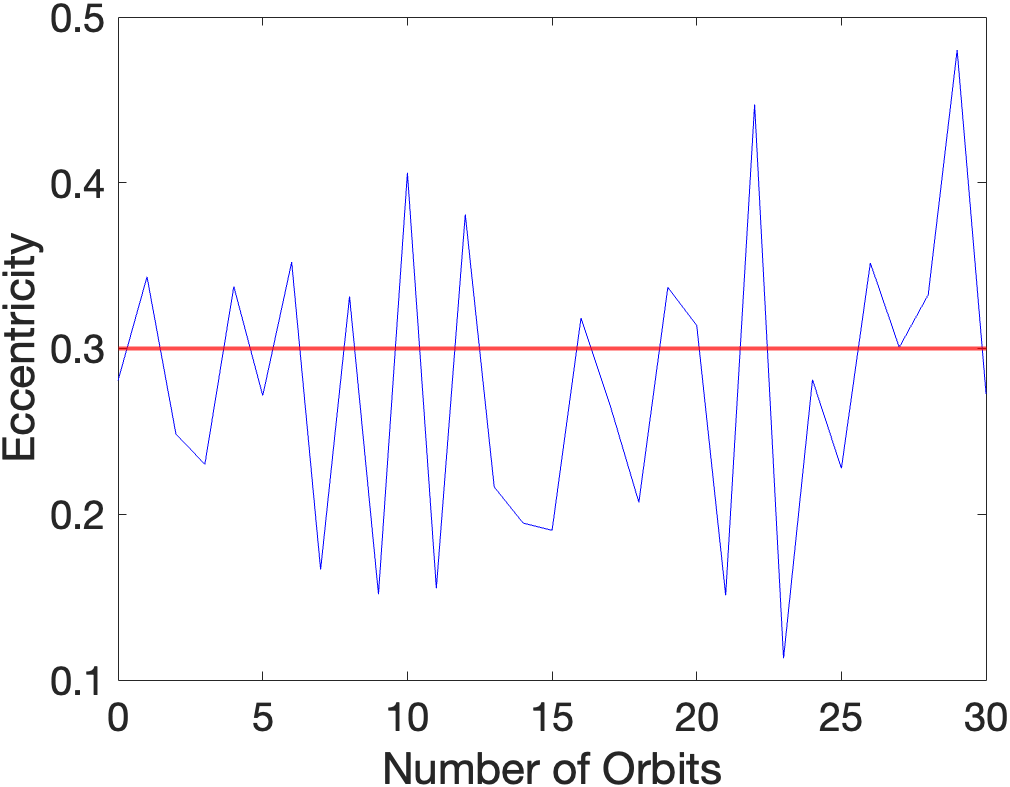}
    \end{subfigure}
    \begin{subfigure}{0.49\textwidth}
        \centering
        \includegraphics[width=\textwidth]{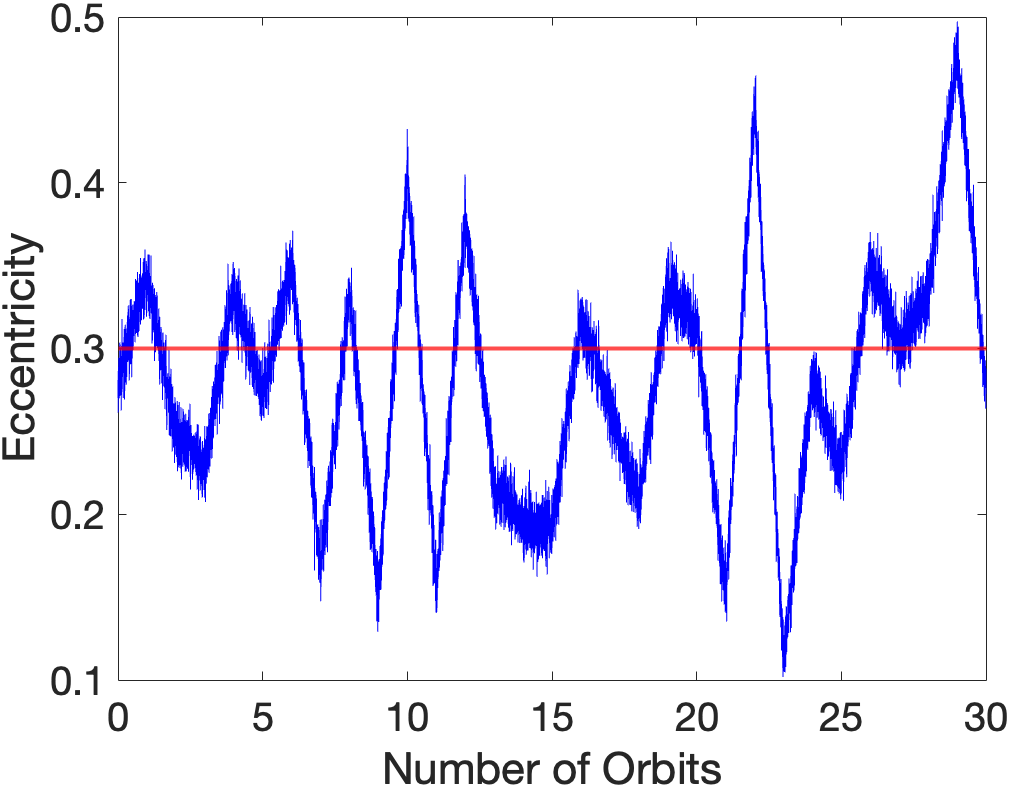}
    \end{subfigure}
    \caption{Two-scale eccentricity model with mean eccentricity $e=0.3$ (red horizontal line) over 30 orbits with $\sigma_\text{orbit}\equiv\sigma_\text{year}=0.1$. The left figure has no daily fluctuations and the right figure has $\sigma_\text{sub-orbit}\equiv\sigma_{\text{day}}=0.01$.  See text for more details regarding the convention for timescales. See also Figures 1, 4, 5 and 7 of \cite{LeungLee2013AnalyticTheory}, showing the interannual and higher frequency eccentricity variations from a factor of 2 to 8 depending on the circumbinary system, and similarly Figure S3 of \cite{Doyle2011Kepler16}.}
    \label{fig:two_scale_eccentricity}
\end{figure}

The simplest means of examining the effect of eccentricity fluctuations described in \ref{se:exo} in the energy balance model is to use a two time-scale eccentricity characterized by two distinct standard deviations: $\sigma_{\text{orbit}}$ for orbital-scale variations and $\sigma_{\text{sub-orbit}}$ for the higher frequency fluctuations, where $\sigma_{\text{sub-orbit}} \ll \sigma_{\text{orbit}}$, as shown in Figure~\ref{fig:two_scale_eccentricity}.  We note, however, that given the potential reader is {\em almost surely} from Earth, for concreteness we ascribe one orbit to one year and the sub-orbit timescale to one day throughout, so that $\sigma_\text{orbit}\equiv\sigma_\text{year}$ and $\sigma_\text{sub-orbit}\equiv\sigma_{\text{day}}$.

At the beginning of each year, the eccentricity is assigned a random value drawn from a distribution with standard deviation $\sigma_{\text{year}}$. If day-to-day fluctuations are neglected ($\sigma_{\text{day}} = 0$), the eccentricity evolves from its value at the beginning of a given year toward the value assigned for the following year. Superimposed on this are small amplitude daily perturbations with variance $\sigma_{\text{day}}$. 
An empirical multiple-scale analysis would tailor the amplitude of the fluctuations according to the time scales discussed in \S \ref{se:intro}, but we reserve that for a separate study.


\section{Results of numerical simulations}\label{se:results}

Here we present representative numerical results of system~\eqref{eq:system_stoch} using \eqref{eq:scheme}, with different choices of the standard deviation $\sigma$ associated with stochastic perturbations of the eccentricity $e$. We examine the bifurcation behavior from a Snowball state to a non-Snowball state. 

We solve the system using the standard forward Euler method as
\begin{equation}
E(t+\Delta t)=E(t)+\Delta t\Bigg[\dfrac{(1-\alpha_i)S_0}{\pi r^2_{\text{noise}}(t)}+Q_{adv}-A-B.T_s(t)\Bigg],
\label{eq:scheme}
\end{equation}
where $r_{\text{noise}}(t)$ is computed numerically from Kepler's equation and a random value of eccentricity $e_{\text{noise}}(t)$ chosen from a normal distribution $N(e,\sigma)$ at time $t$. Note that the advective flux $Q_{adv}$ includes $e_{\text{noise}}$, and thus it can also have a significant stochastic effect on the evolution equation depending on the magnitude of the variance $\sigma^2$.

\subsection{Sensitivity to eccentricity fluctuations}

\begin{figure}[H]
    \centering
    \includegraphics[width=\linewidth]{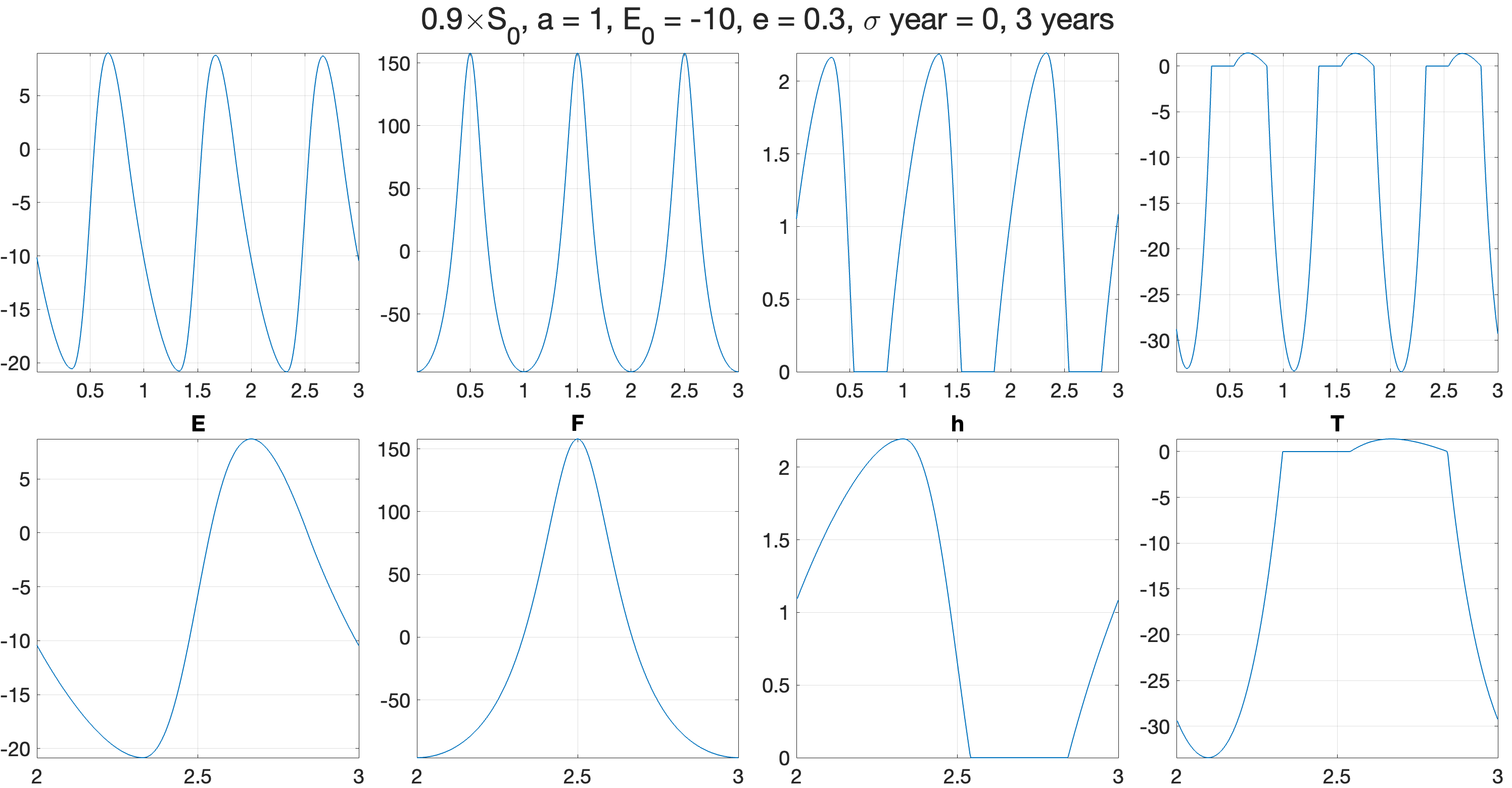}
    \caption{The deterministic solution converges to a stable solution in 3 years. The top panels show $E,\,F,\,h_i,\,T_s$ for all years, and the bottom panels show the last year, when $e=0.3$ and $0.9\times S_0$}
    \label{fig:e_03_deter}
\end{figure}

\begin{figure}[H]
    \centering
    \includegraphics[width=\linewidth]{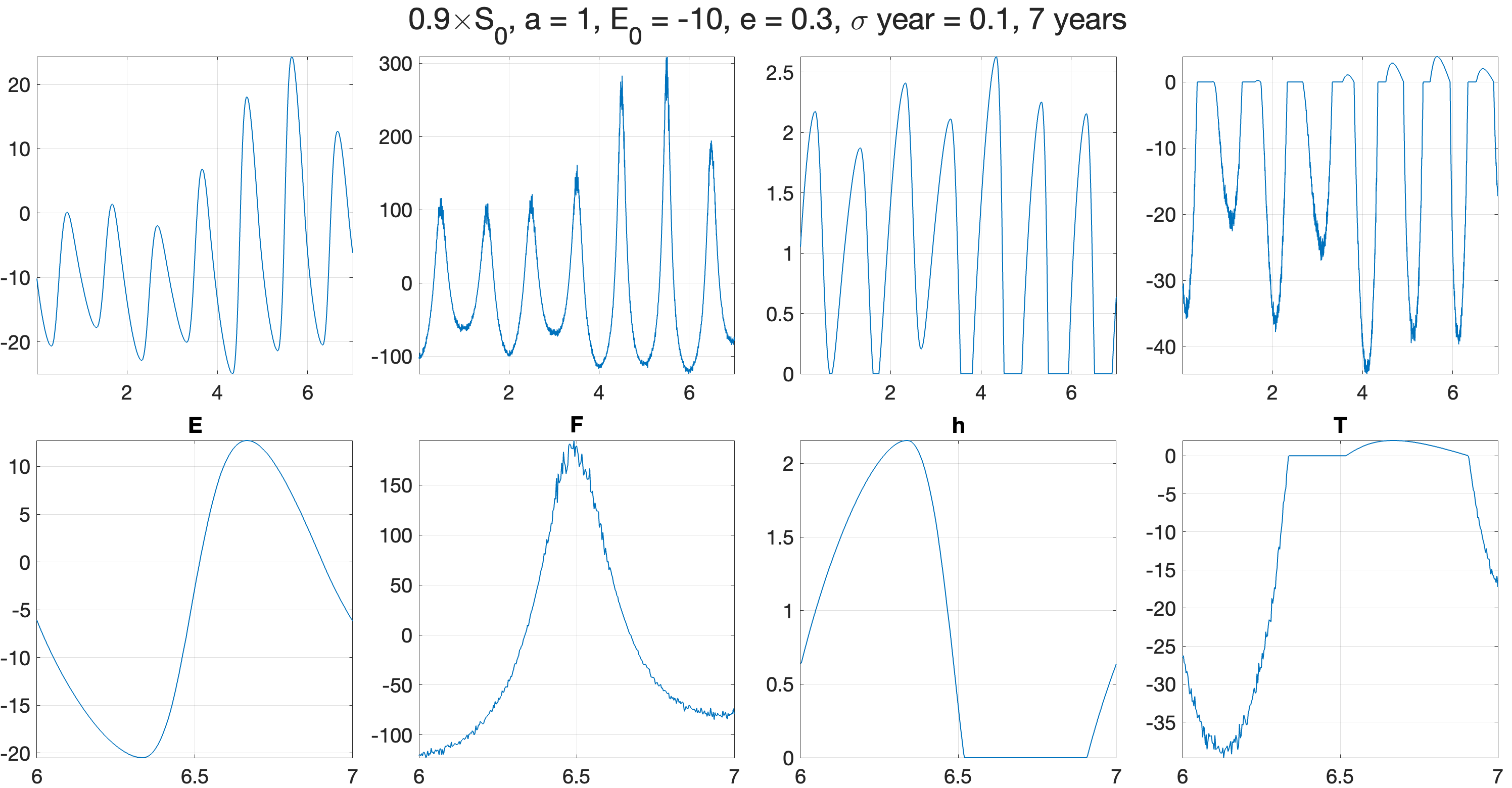}
    \caption{Same as Fig. \ref{fig:e_03_deter}, but for the stochastic solution with $\sigma_{year}=0.1$, which converges to a stable solution in 7 years. The top row is for all years, and the bottom row is for the last year.}
    \label{fig:e_03_stochastic_1}
\end{figure}

\begin{figure}[H]
    \centering
    \includegraphics[width=\linewidth]{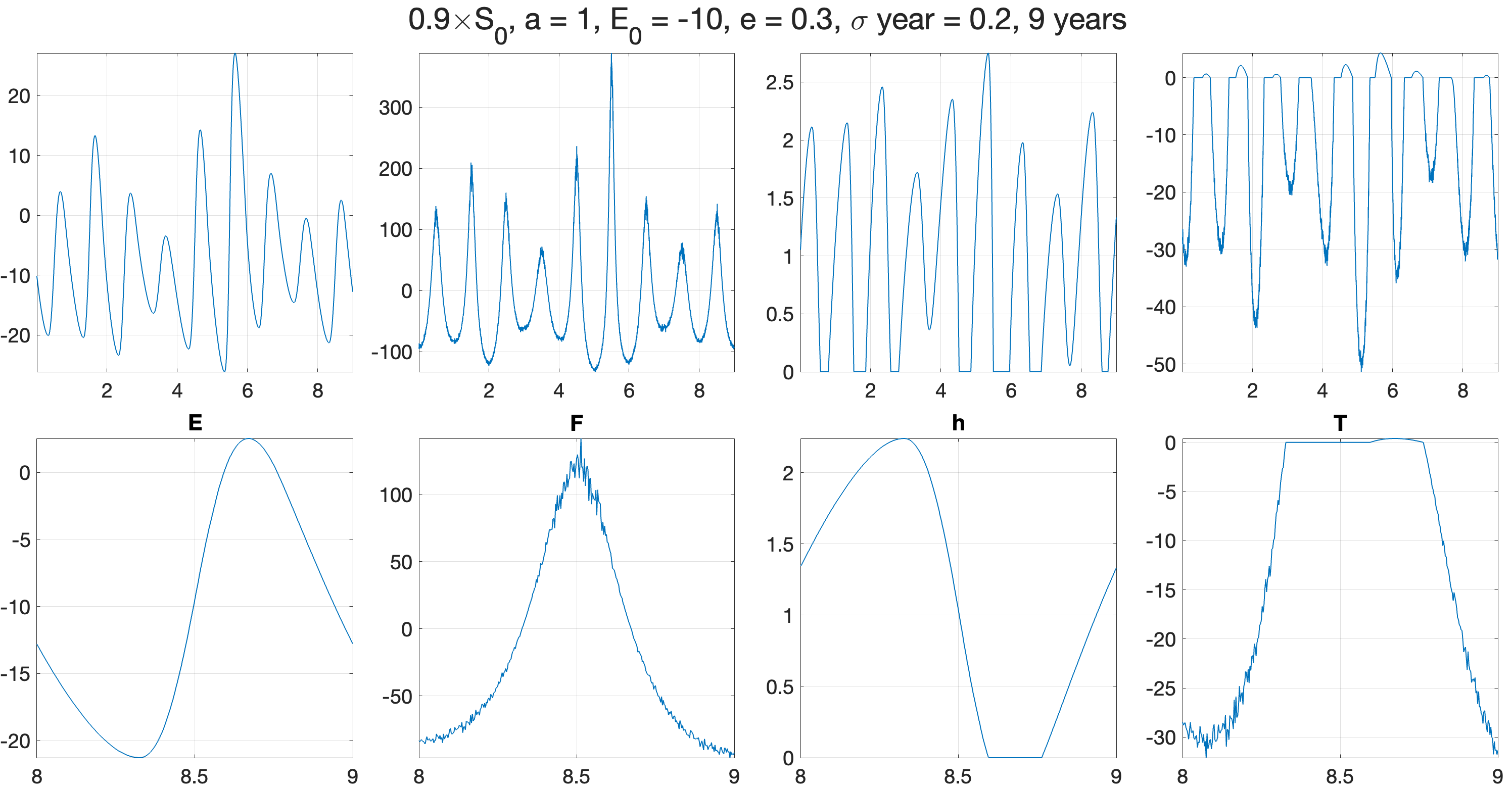}
    \caption{Same as Figs. \ref{fig:e_03_deter},\ref{fig:e_03_stochastic_1}, but for the stochastic solution with $\sigma_{year}=0.2$, which converges to a stable solution in 9 years. The top row is for all years, and the bottom row is for the last year.}\label{fig:e_03_stochastic_2}
\end{figure}

We study this system with eccentricity $e=0.3$ and stellar flux $0.9\times S_0$. We set the semi-major axis to $a=1$ AU and initiate the model with $E_0=-10$. One year is fixed at 360 days, so the time step is $\Delta t=1/360$ year. 
We set the standard deviation of daily eccentricity to $\sigma_{day}=\min(10\%\sigma_{year};\,0.01)$, whose maximum value is limited to 0.01, and restrict our study to bifurcations that occur within 10,000 years. The top panels of Figure \ref{fig:e_03_deter} show the deterministic solutions for $E,\,F,\,h_i,\,T_s$ over the time it takes to reach a stable seasonal cycle, and the bottom panels show the last year of the process. Figures \ref{fig:e_03_stochastic_1}-\ref{fig:e_03_stochastic_2} show the same but for two stochastic solutions with standard deviations for perturbed eccentricity of $\sigma_{year}=0.1$ and $\sigma_{year}=0.2$ respectively. Comparing Figures \ref{fig:e_03_deter}-\ref{fig:e_03_stochastic_2} clearly shows that the role of fluctuations in eccentricity is to amplify the asymmetry and intermittency in radiative flux forcing, thereby amplifying the time it takes to reach a stable seasonal cycle from 3 to 9 years as $\sigma_{year}$ increases from 0 to 0.2.


\subsection{Sensitivity to mean annual stellar flux and eccentricity fluctuations}

\begin{figure}[H]
    \centering
    \begin{subfigure}{0.49\textwidth}
        \centering
        \includegraphics[width=\textwidth]{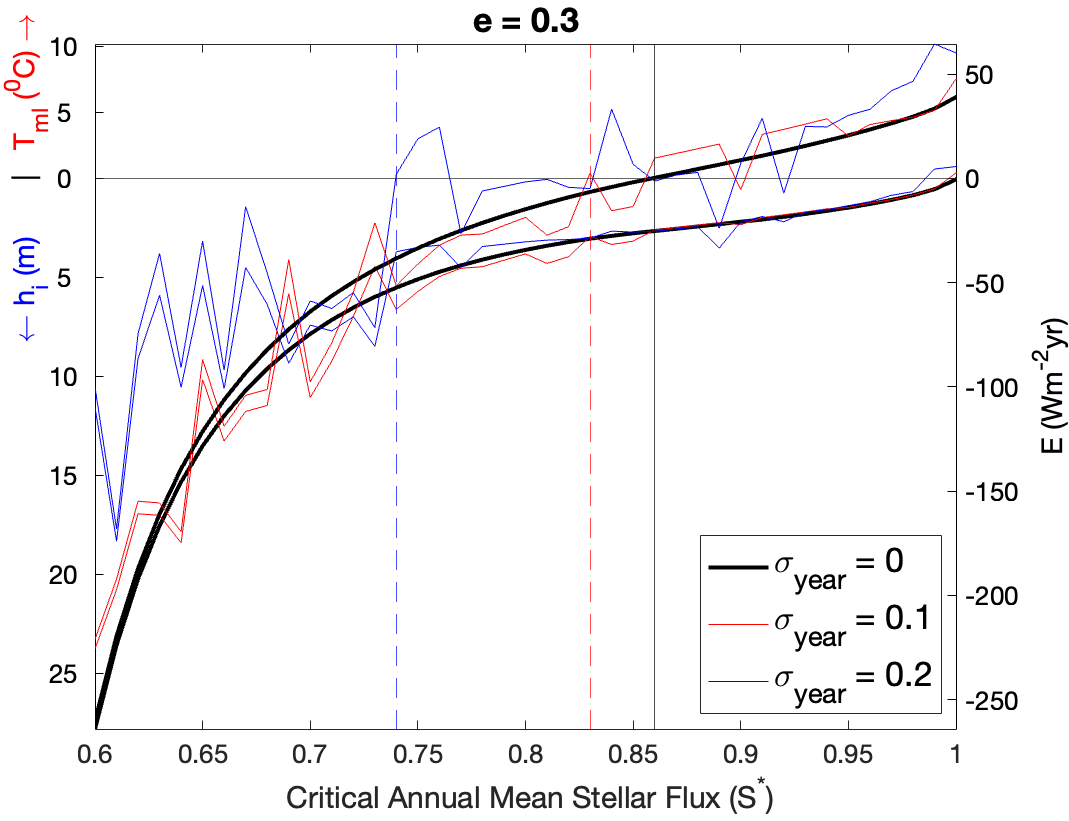}
    \end{subfigure}
    \begin{subfigure}{0.49\textwidth}
        \centering
        \includegraphics[width=\textwidth]{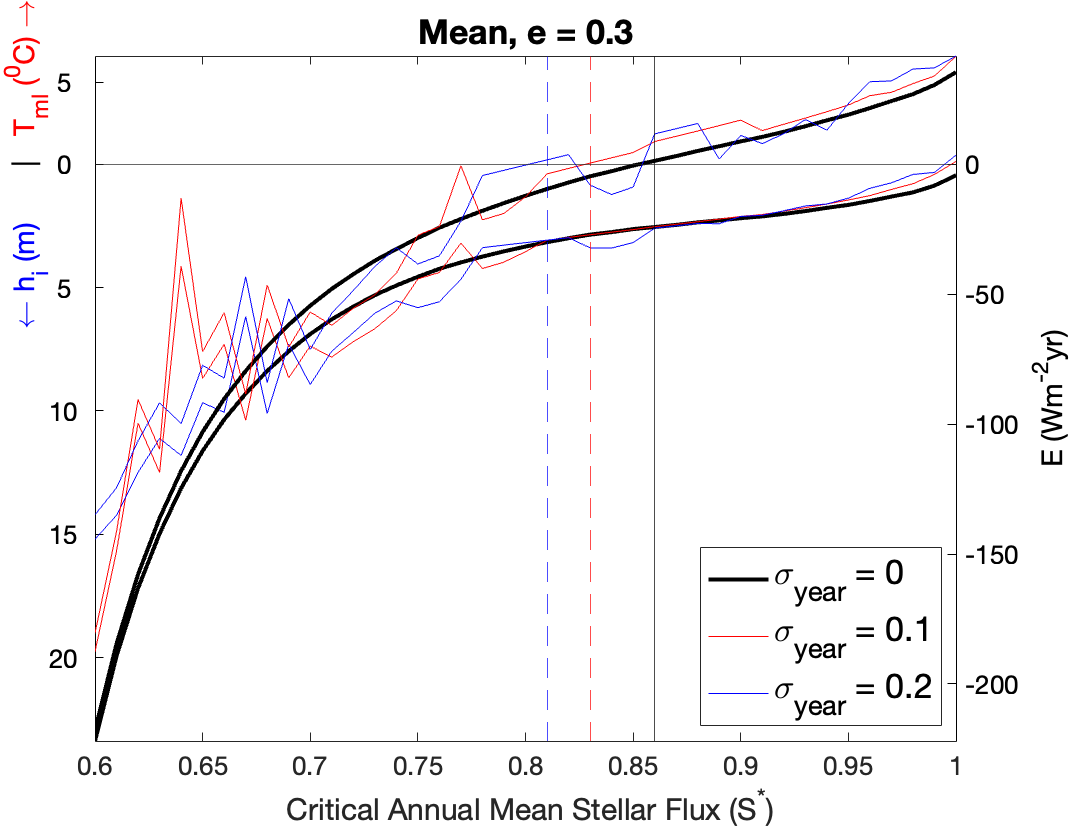}
    \end{subfigure}
    \caption{Bifurcation diagram with mean eccentricity $e=0.3$: deterministic system ($\sigma_{year}=0$) and stochastic system with eccentricity noise characterized by $\sigma_{year}=0.1$ (red) and $\sigma_{year}=0.2$ (blue). Vertical lines indicate the critical annual mean stellar flux $S^*$ where the system transitions from a Snowball to a non-Snowball state. Depending on the sign of $E$, all these values are converted to the ice thickness or the ocean mixed-layer temperature. \textbf{\textit{Left}}: The upper lines are for the maximum values of $E$, and the lower lines are for the corresponding minimum values of $E$ in the convergence year. \textbf{\textit{Right}}: The upper lines are for the mean maximum values of $E$, and the lower lines are for the corresponding mean minimum values of $E$ through all the years up to the year of convergence.}
    \label{fig:bifur_fixed_e_03}
\end{figure}

First, we calculate the critical mean annual stellar flux $S^*$ associated with the bifurcation. Figure \ref{fig:bifur_fixed_e_03} shows three cases: deterministic ($\sigma_{year}=0$) and stochastic, with $\sigma_{year}=0.1$ and $\sigma_{year}=0.2$, all with fixed mean eccentricity $e=0.3$. In the deterministic case, for each $S^*$ if the system converges to a stable solution, we use the maximum and minimum values of $E$ in the last year of simulations, and then, depending on the sign of $E$, plot the converged ice thickness or ocean mixed-layer temperature.  The two principal results of this figure are that, as expected, the role of eccentricity fluctuations is to ``smear out'' the bifurcation to the non-Snowball state, and that as the magnitude of these fluctuations increases the bifurcation occurs at smaller critical mean annual stellar fluxes $S^*$.  This is consistent with the notion of a bifurcation point transitioning to a bifurcation region in a noisy system \citep{MeunierVerga1988}.  Although the non-autonomous deterministic backbone of the present formulation in Eq.~\eqref{eq:system_main} is not a canonical autonomous deterministic normal form, the fact that the nature of the noise (additive, multiplicative or colored) in such circumstances can change the nature of the bifurcation itself \citep{KnoblochWiesenfeld1983}, is of interest in further studies of our system.  


\begin{figure}[H]
    \centering
    \begin{subfigure}{0.49\textwidth}
        \centering
        \includegraphics[width=\textwidth]{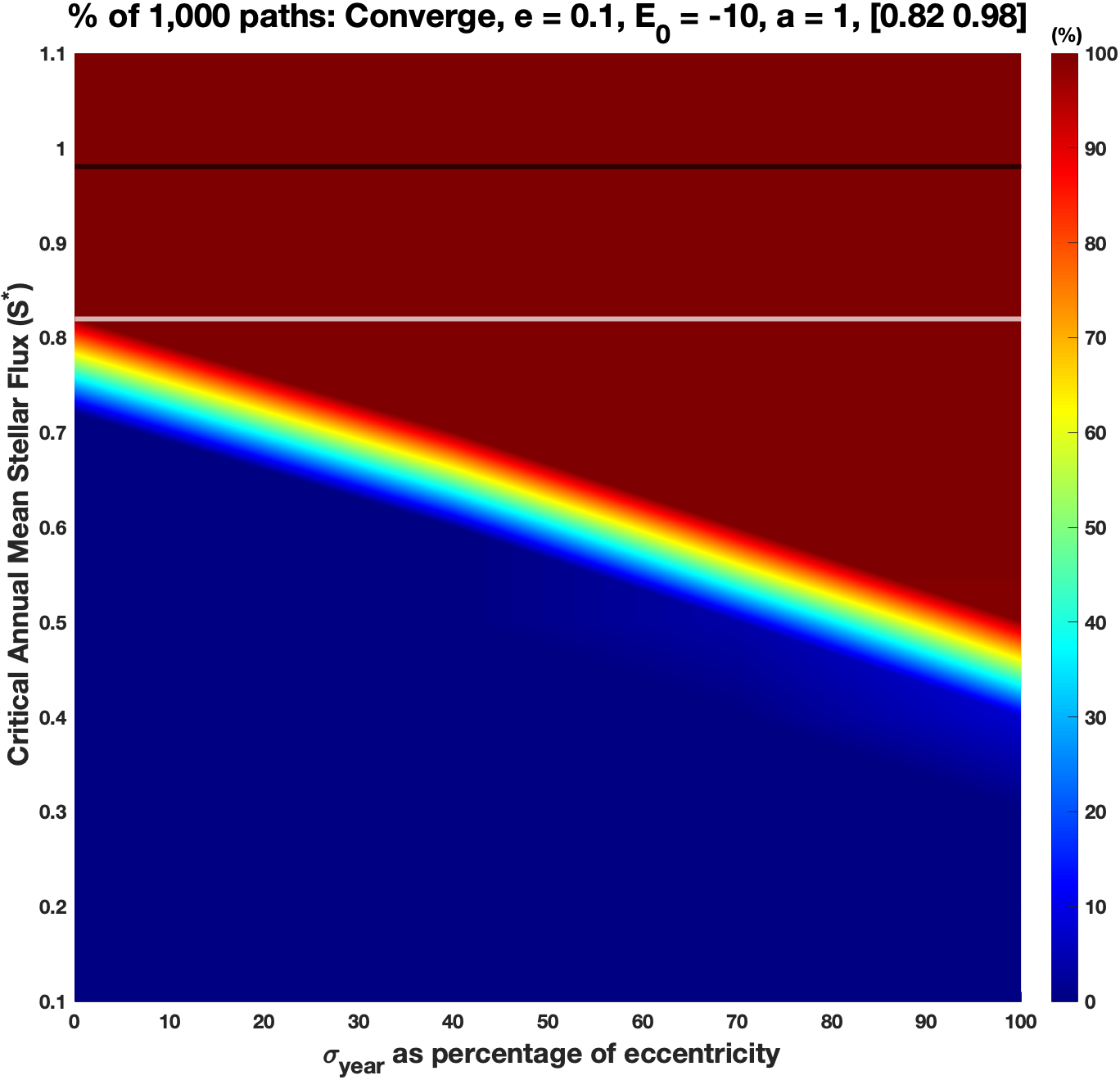}
    \end{subfigure}
    \begin{subfigure}{0.49\textwidth}
        \centering
        \includegraphics[width=\textwidth]{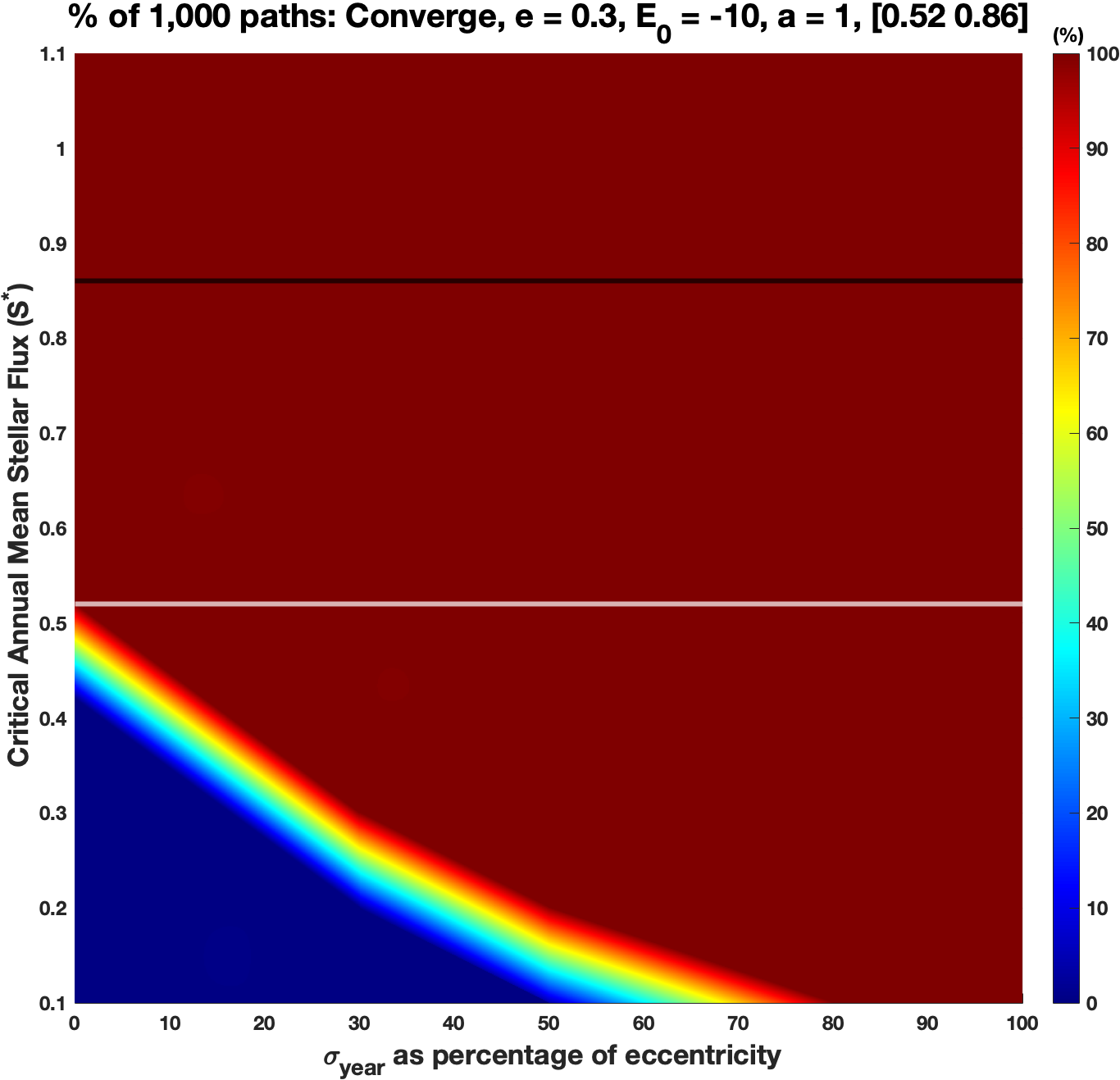}
    \end{subfigure}
    \caption{Percentage of 1,000 stochastic paths that converge shown as color maps for $e=0.1,\,0.3$. The two bracketed values in the title of each sub-figure are the critical deterministic values of $S^*$ for convergence and bifurcation given in Table \ref{tab:three_critical_Sstar}. 
    The white and black horizontal lines are the convergence and bifurcation values respectively. Additional figures are found in the Supplementary Material (\ref{sec:SM}).}
    \label{fig:percent_con_1000paths}
\end{figure}

\begin{figure}[H]
    \centering
    \begin{subfigure}{0.49\textwidth}
        \centering
        \includegraphics[width=\textwidth]{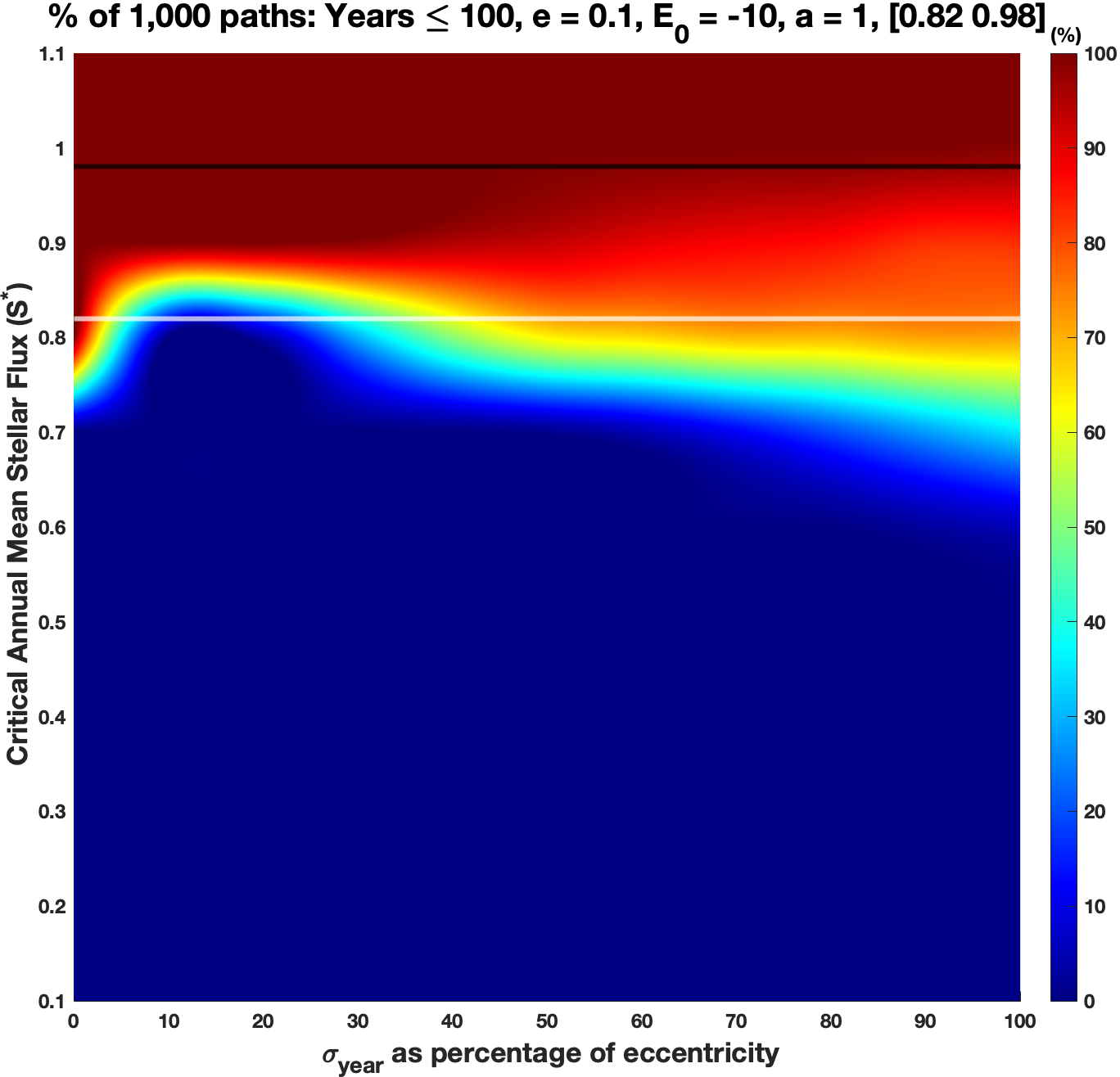}
    \end{subfigure}
    \begin{subfigure}{0.49\textwidth}
        \centering
        \includegraphics[width=\textwidth]{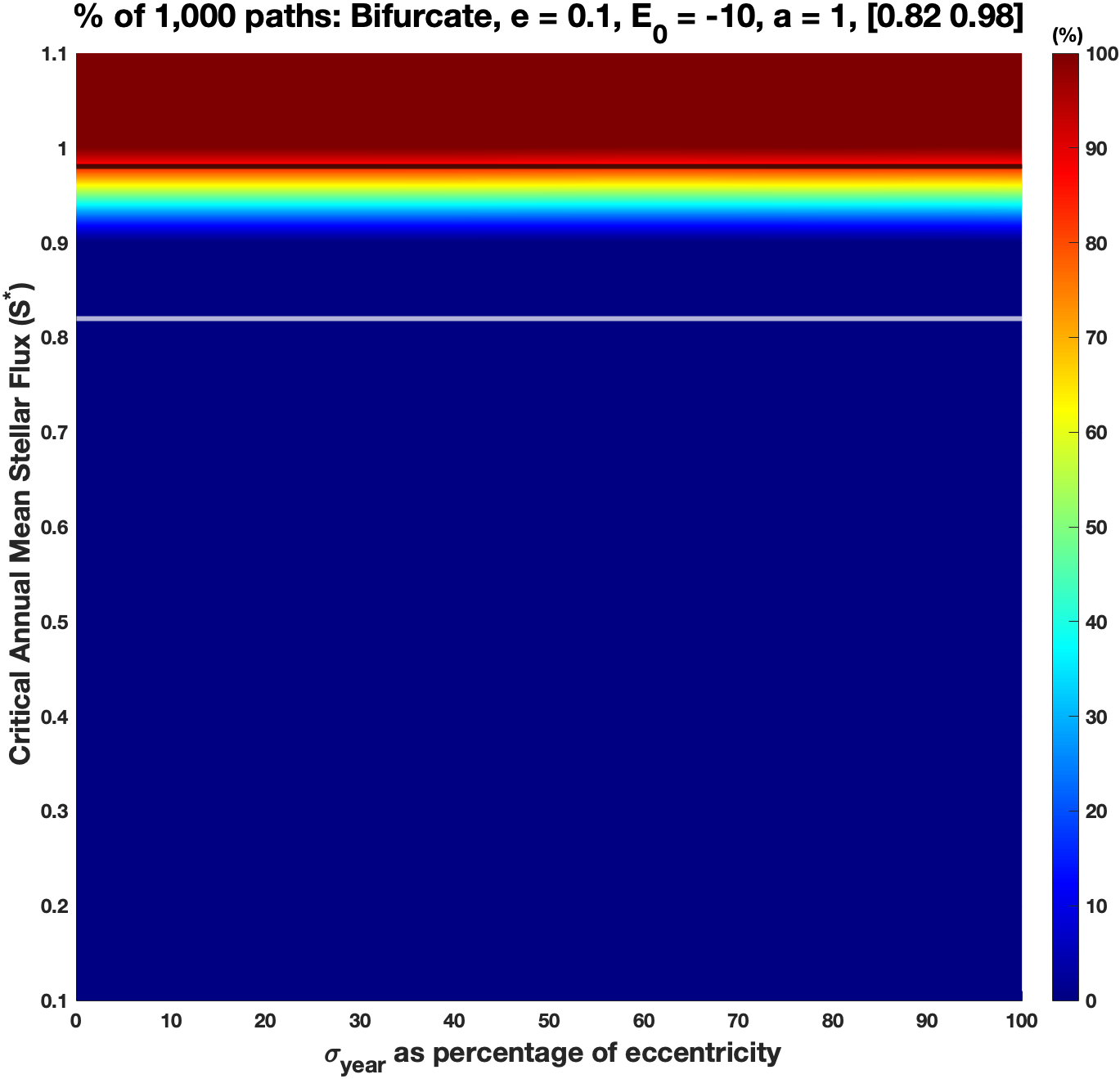}
    \end{subfigure}

    \begin{subfigure}{0.49\textwidth}
        \centering
        \includegraphics[width=\textwidth]{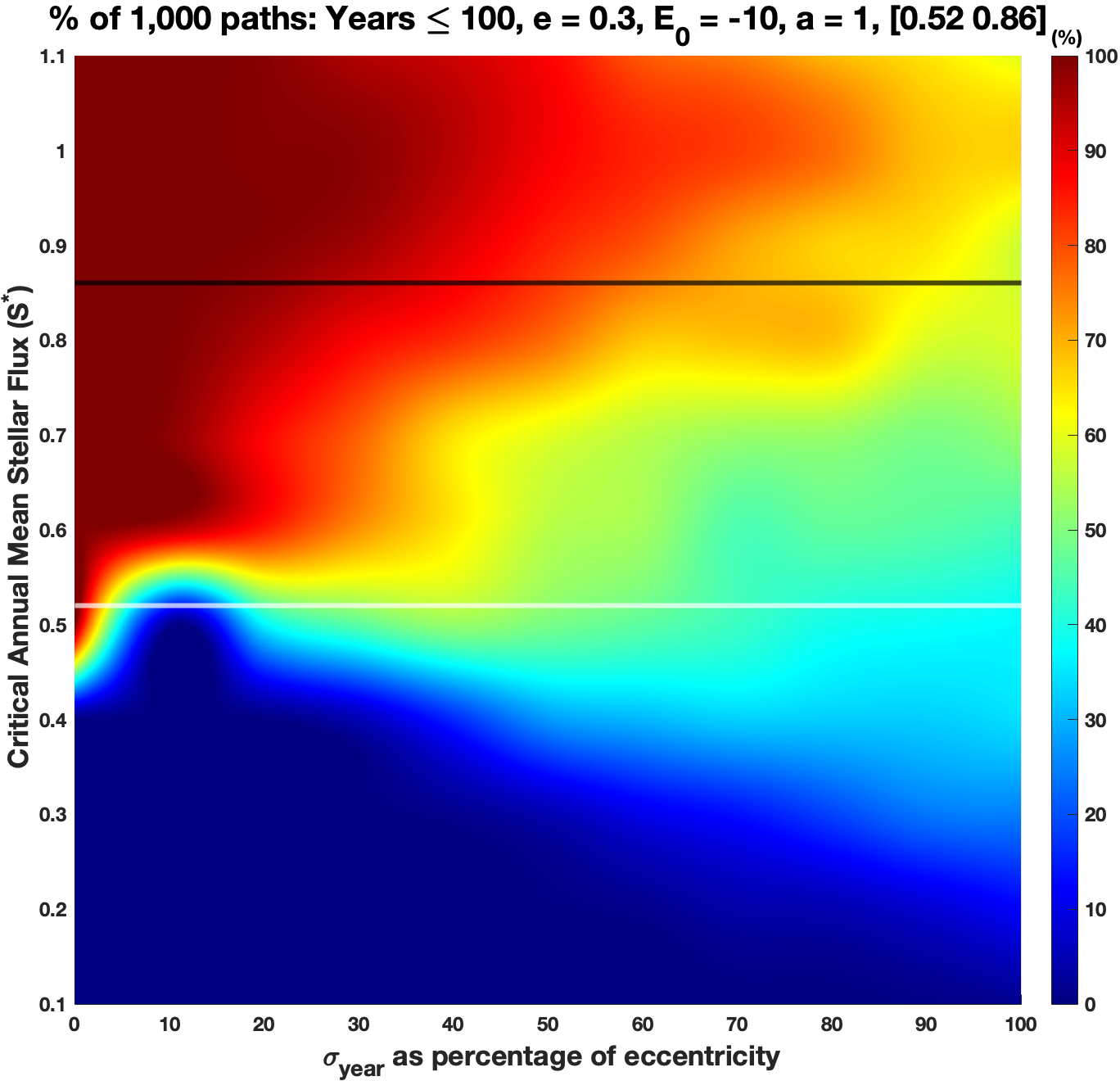}
    \end{subfigure}
    \begin{subfigure}{0.49\textwidth}
        \centering
        \includegraphics[width=\textwidth]{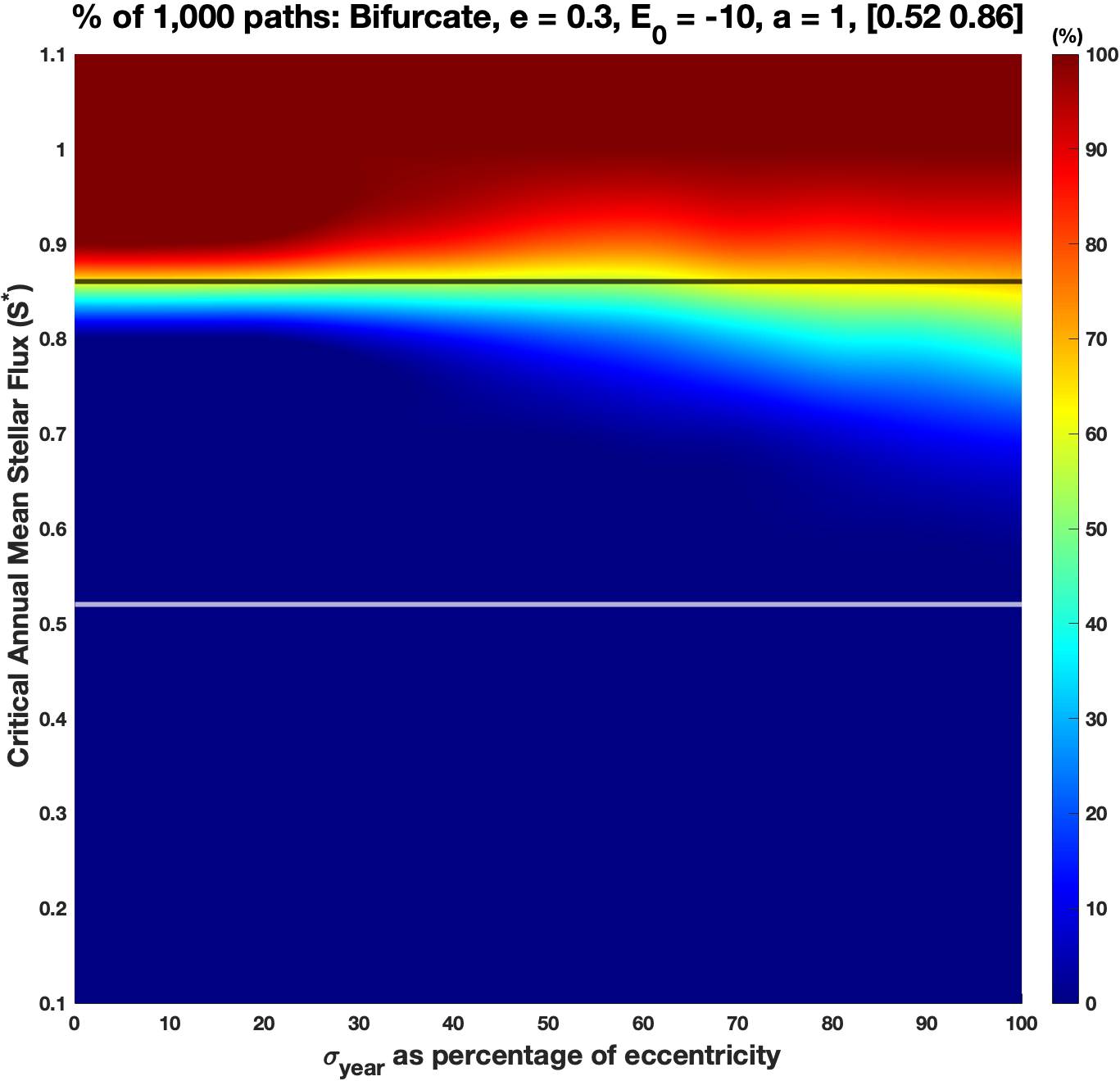}
    \end{subfigure}

    \begin{subfigure}{0.49\textwidth}
        \centering
        \includegraphics[width=\textwidth]{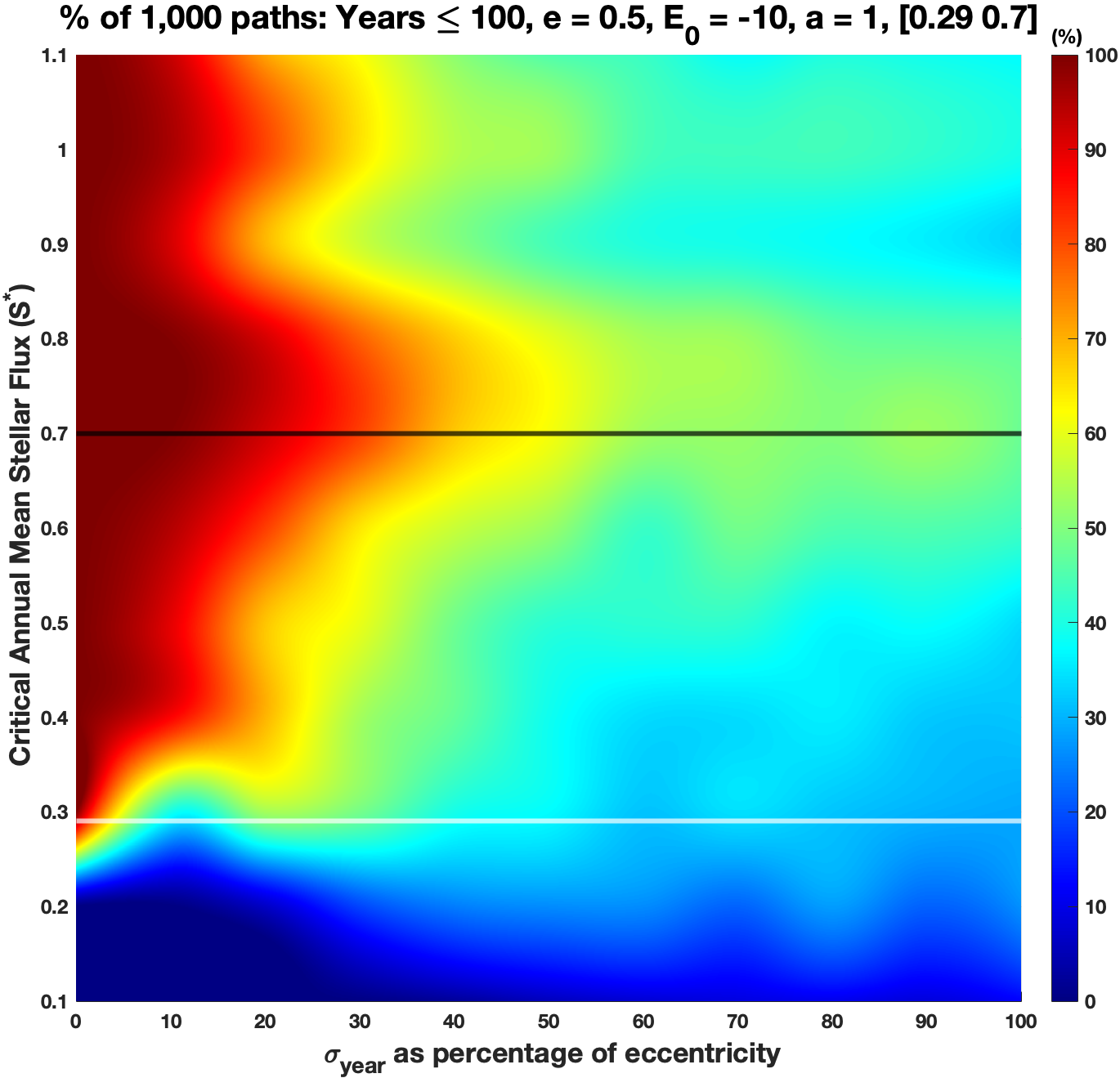}
    \end{subfigure}
    \begin{subfigure}{0.49\textwidth}
        \centering
        \includegraphics[width=\textwidth]{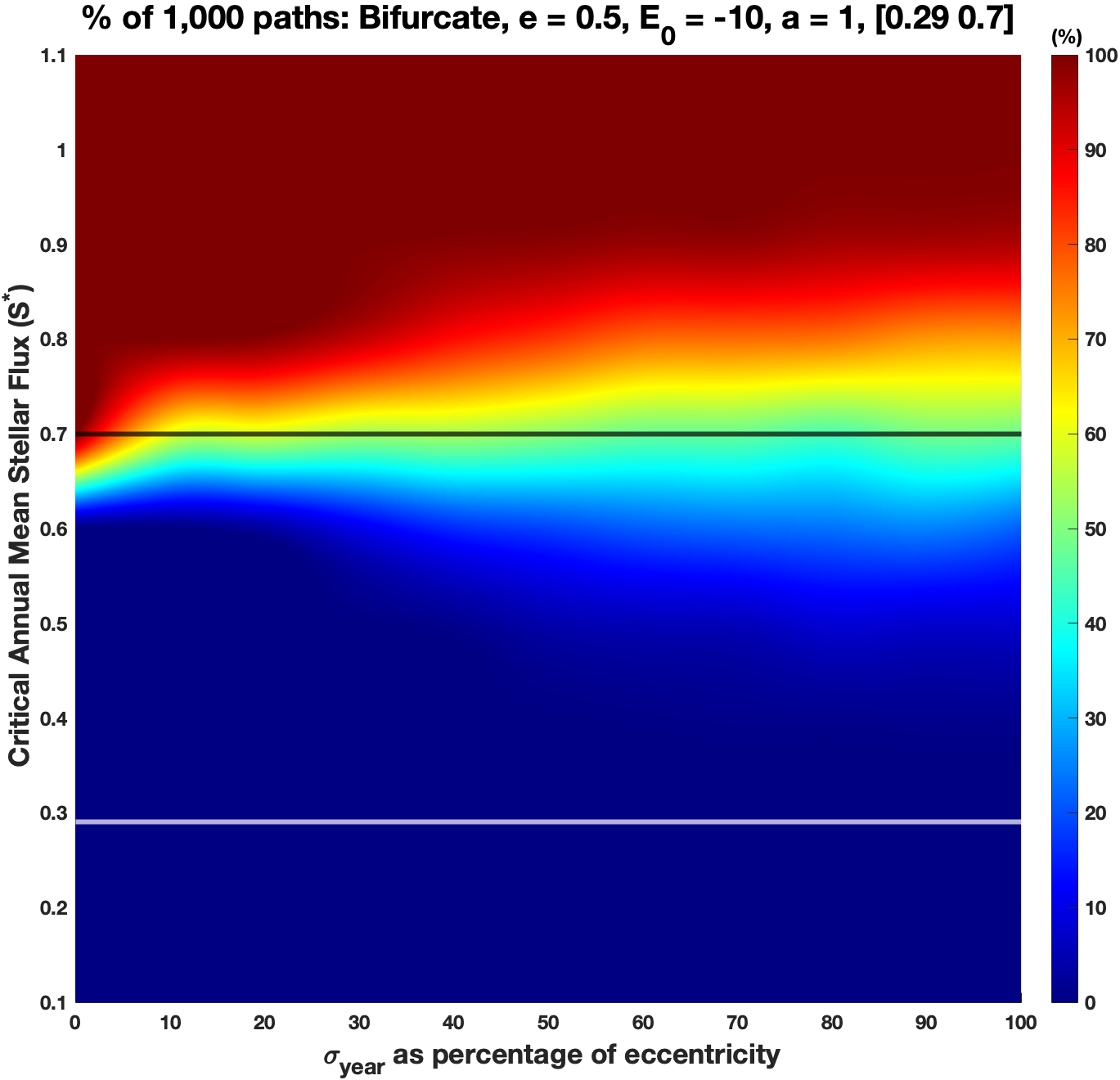}
    \end{subfigure}
    \caption{Percentage of 1,000 stochastic paths that \textbf{\textit{Left:}} converge and \textbf{\textit{Right:}} bifurcate from the Snowball state to the non-Snowball state shown as color maps for $e=0.1,\,0.3,\,0.5$. The two bracketed values in the title of each sub-figure are the critical deterministic values of $S^*$ for convergence and bifurcation given in Table \ref{tab:three_critical_Sstar}. 
    The white and black horizontal lines are the convergence and bifurcation values respectively. Additional figures are found in the Supplementary Material (\ref{sec:SM}).}
    \label{fig:percent_yr_bif_1000paths}
\end{figure}

Second, we investigate the dependence of the convergence to a steady solution and the bifurcation from the Snowball state to the non-Snowball state of the stochastic system on $S^*$ and the mean and fluctuations in eccentricity. We run the simulations for six values of eccentricity from $e=0.1$ to 0.6 in increments of 0.1. The values in Table \ref{tab:three_critical_Sstar} are plotted as horizontal lines in Figures \ref{fig:percent_con_1000paths}, \ref{fig:percent_yr_bif_1000paths}, where the white line is the critical convergence value and the black line is the deterministic bifurcation point. The horizontal axis is the $\sigma_{year}$ as a percentage of the mean eccentricity $e$ and the vertical axis is $S^*$.
For each pair of $S^*$ and $\sigma_{year}$, we run the stochastic system 1,000 times or 1,000 paths of a stochastic chain of eccentricity. Then we count the number of paths for which the system converges and bifurcates and plot these as a percentage color map on the right of each sub-figure.

As shown in Figures \ref{fig:e_03_deter}-\ref{fig:e_03_stochastic_2}, for a fixed mean eccentricity, as $\sigma_{year}$ increases, so too does the time to convergence due to the asymmetry and intermittency in radiative flux forcing.
Figure \ref{fig:percent_con_1000paths} shows that for a fixed mean eccentricity, the percentage of the stochastic trajectories that converge to the minimal deterministic critical annual mean stellar flux---the white horizontal line---increases with $\sigma_{year}$.  Moreover, because $S^*$ decreases as mean eccentricity increases, the $S^*$--$~\sigma_{year}$ convergence domain also increases.  These effects are combined in Figure \ref{fig:percent_yr_bif_1000paths} (left), where we plot the percentage of 1,000 stochastic trajectories that converge within 100 years. Clearly, as both the mean and the variance of eccentricity fluctuations increase, convergence to a stable steady state within a century becomes less probable due to the stochasticity in the radiative flux forcing shown in Figures \ref{fig:e_03_deter}-\ref{fig:e_03_stochastic_2}.  Importantly, this effect provides a basic bound on the existence of the Snowball to non-Snowball bifurcation, which implies that there may be many high eccentricity planetary systems in a perpetually transient state. 

Figure \ref{fig:percent_yr_bif_1000paths} (right) shows that for small mean eccentricity the Snowball to non-Snowball bifurcation of the stochastic system approaches that of the deterministic system, given by the black horizontal line.  Although as the mean eccentricity increases, the deterministic bifurcation threshold $S^*$ decreases, as the eccentricity fluctuations increase the probability of  bifurcations also decreases, consistent with notion of an abrupt bifurcation transitioning to a region in a noisy system \citep{MeunierVerga1988}. See also Figures 9-13 of \citet{MoonWettlaufer2017StochasticSeaIce} for a similar effect shown over the seasonal cycle.


In summary, Figures \ref{fig:percent_con_1000paths}-\ref{fig:percent_yr_bif_1000paths} demonstrate that (1) increasing the mean orbital eccentricity reduces the deterministic critical annual mean stellar flux $S^*$ for the bifurcation from the Snowball to non-Snowball state.  This is consistent with \citet{JiAbbot25}; (2) the effects of stochastic eccentricity on the bifurcation become negligible for mean eccentricities below 0.1, and ``smear out'' the transition (i.e., the abrupt deterministic transition in $S^*$ widens over a range of $\Delta S^*$) as the mean eccentricity increases; (3) for any mean eccentricity as the magnitude of the variance of the eccentricity, $\sigma_{year}$, increases the probability of convergence to a steady state within 100 years in 1000 realizations decreases and the bifurcation region $\Delta S^*$ broadens, such that for many parameter values steady states and bifurcations will only occur rarely.  

\section{Conclusion}\label{se:conclu}

Motivated by the general problem of the role of orbital parameters on planetary climate, and the recent finding of \citet{JiAbbot25} that at moderate eccentricity (0.2--0.3) the planetary Snowball bistability vanishes, we have introduced a minimal energy balance model that includes stochasticity in orbital eccentricity.  Because \citet{JiAbbot25} found a favorable comparison between a GCM and the energy balance model of \citet{EW09}, we introduced stochastic eccentricity into the latter through an analysis of the Kepler equation, as detailed in Section \ref{se:noise} and \ref{se:small_variance}, leading to Eq. \eqref{eq:system_stoch}. While we noted the frequency range of eccentricity fluctuations in the solar system \citep{Laskar2011, Laskar21, Scafetta19}, we chose the simplest treatment: a two time-scale eccentricity characterized by annual and daily variations with distinct standard deviations $\sigma_{\text{day}} \ll \sigma_{\text{year}}$.

In a suite of numerical studies of this model we find that, apart from mean eccentricities less than 0.1, the stochastic dynamics are quantitatively and qualitatively different than the deterministic dynamics.  We find that for a fixed mean eccentricity, fluctuations in eccentricity amplify the asymmetry and intermittency in radiative flux forcing, thereby increasing the time it takes to reach a stable seasonal cycle (See e.g. Figures \ref{fig:e_03_deter}-\ref{fig:e_03_stochastic_2}).  As the mean eccentricity increases above 0.1, the stochastic transition from the Snowball to non-Snowball state occurs over a range of mean stellar flux $\Delta S^*$ rather that at a particular stellar flux $S^*$ for the deterministic case. Moreover, for any mean eccentricity $\ge 0.1$, as $\sigma_{year}$ increases, $\Delta S^*$ increases and the probability of the system reaching a steady state and experiencing a bifurcation decreases.
Importantly, this implies that there may be many high eccentricity planetary systems in a perpetually transient state. 


\section*{Acknowledgements}

The authors gratefully acknowledge comments by Dorian Abbot and support from Yale University and the Swedish Research Council under Grant No.638-2013-9243.

\bibliographystyle{apalike}
\bibliography{Reference}




\appendix

\section{Critical values of $S^*$ for energy convergence and bifurcation thresholds}\label{se:E_decrease}

For initial values of $E<0$ and $F<0$, we rewrite equation \eqref{eq:system_main} as
\begin{equation}
\dfrac{dE}{dt}=F(t)-F(t)\bigg(1+\dfrac{k_i}{Bh_i}\bigg)^{-1}=F(t)\Bigg[1-\bigg(1+\dfrac{k_i}{Bh_i}\bigg)^{-1}\Bigg].
\end{equation}
Since $1+\dfrac{k_i}{Bh_i}>1$, then $1-\bigg(1+\dfrac{k_i}{Bh_i}\bigg)^{-1}>0$. Thus, if $F(t)<0$ for all $t$ in the orbit, then $\dfrac{dE}{dt}<0$ and hence $E$ will decrease without bound, so the system will not converge. From the formula for $F(t)$, we conclude that the maximum value of $F(t)$ must be negative. This occurs when $r$ reaches the minimum value, $r=a(1-e)$, from which we have
\begin{equation}\label{eq:Sstar_critical_converge}
\dfrac{(1-\alpha_i)S^*S_0}{\pi a^2(1-e)^2}+Q_{adv}-A<0\Leftrightarrow S^*<S^*(e)=\dfrac{(A-Q_{adv})\pi a^2(1-e)^2}{(1-\alpha_i)S_0}.
\end{equation}
We plot the function $S^*(e)$ in Figure \ref{fig:conv_bif} as the blue curve. To verify this critical transition with numerical simulations, for each value of eccentricity $e=0.1,\,0.2,\,...,0.6$, we numerically run the system for many values of $S^*=0.01,\,0.02,\,...,1.1$ to find the first point at which the system converges within 10,000 years, which are the blue points in Figure \ref{fig:conv_bif}. The numerical bifurcation values are the red points obtained similarly. Table \ref{tab:three_critical_Sstar} lists these values for each eccentricity $e$.

\begin{minipage}{\textwidth}
  \begin{minipage}[b]{0.54\textwidth}
    \centering
    \includegraphics[width=1.05\textwidth]{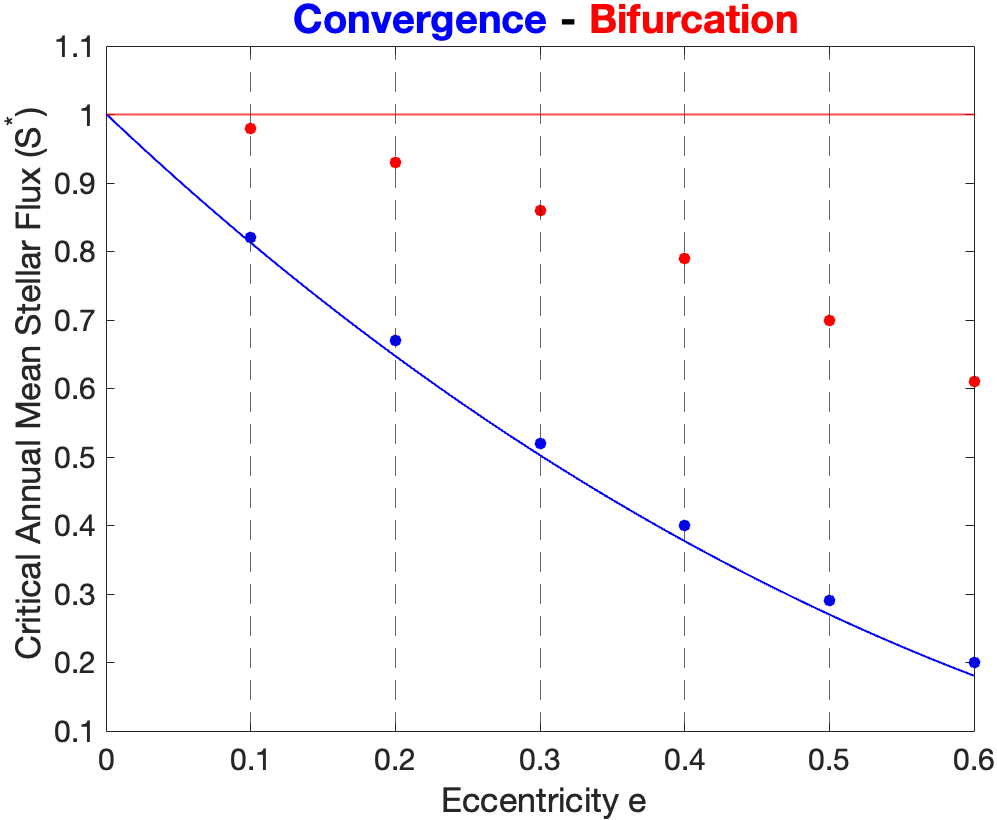}
    \captionof{figure}{Deterministic convergence and bifurcation thresholds.}
    \label{fig:conv_bif}
  \end{minipage}
  \hfill
  \begin{minipage}[b]{0.45\textwidth}
    \centering
    \begin{table}[H]
\begin{center}
\begin{tabular}{ \thickvertical c|c|c \thickvertical} 
\thickhline
$e$ & Convergence & Bifurcation\\
 \thickhline
0.1 & 0.82 & 0.98\\
0.2 & 0.67 & 0.93\\
0.3 & 0.52 & 0.86\\
0.4 & 0.40 & 0.79\\
0.5 & 0.29 & 0.70\\
0.6 & 0.20 & 0.61\\
\thickhline
\end{tabular}
\end{center}
\end{table}
      \captionof{table}{Critical values of $S^*$}\label{tab:three_critical_Sstar}
    \end{minipage}
  \end{minipage}

\begin{figure}[H]
    \centering
    \includegraphics[width=\linewidth]{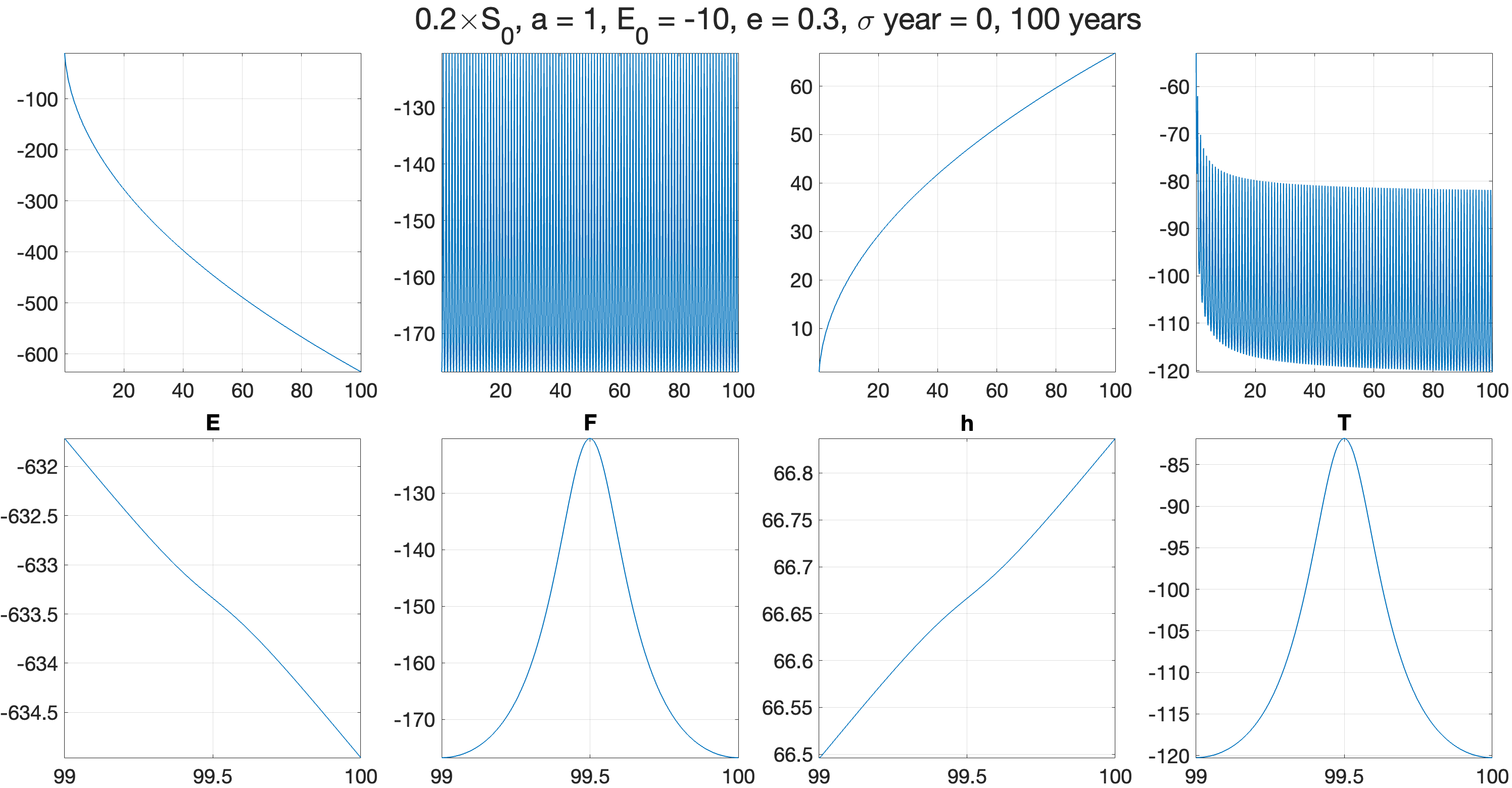}
    \caption{$e=0.3$ and $0.2\times S_0$: The deterministic system does not converge to a stable solution in 100 years. The top row is for all years, and the bottom row is for the last year.}
    \label{fig:nonconverge_sigma0}
\end{figure}
\begin{figure}[H]
    \centering
    \includegraphics[width=\linewidth]{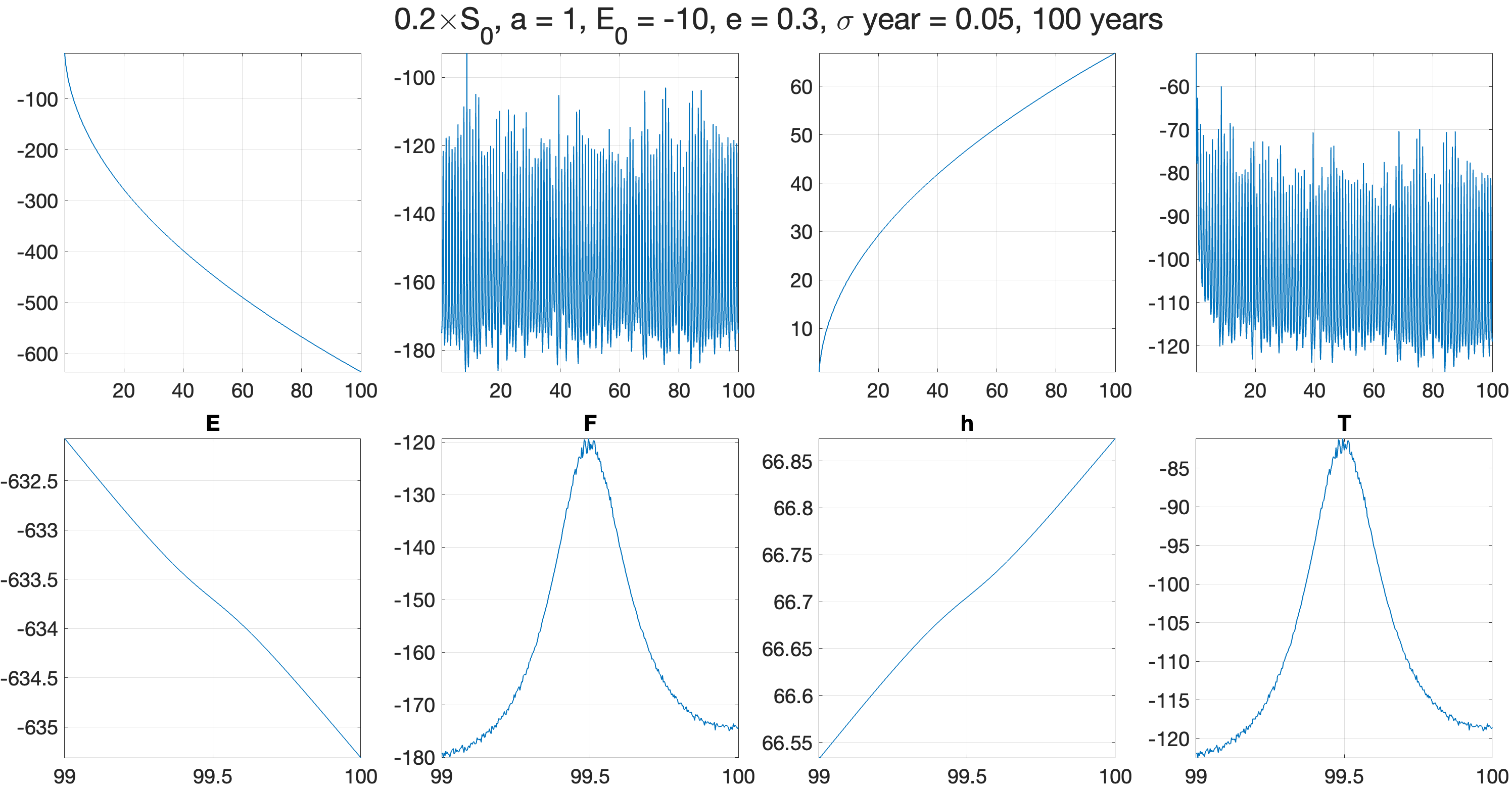}
    \caption{$e=0.3$ and $0.2\times S_0$: The stochastic system with $\sigma_{year}=0.05$ does not converge to a stable solution in 100 years. The top row is for all years, and the bottom row is for the last year.}
    \label{fig:nonconverge_sigma005}
\end{figure}

As described in Section \ref{se:results}, we solve the deterministic system with eccentricity value $e=0.3$, but here with stellar flux $0.2\times S_0$. We set the semi-major $a=1$ AU and initiate the model with $E_0=-10$. One year is 360 days, so the time step is $\Delta t=1/360$ year. In Figure \ref{fig:nonconverge_sigma0}, we show the deterministic solution of the system within 100 years, including $E,\,F,\,h_i,\,T_s$ through all years and the last year of the process. We observe $E$ to decrease without bound and hence the ice grows without bound leading to a non-convergence to a stable solution of the system, or a runaway Snowball state.
In Figure \ref{fig:nonconverge_sigma005} we show the stochastic system with a small standard deviation $\sigma_{year}=0.05$, which also leads to a runaway Snowball state.


\section{Analysis of fluctuations in eccentricity and Kepler's equation}\label{se:small_variance}

We expand the eccentric anomaly $\psi$, in a Fourier series with respect to $M$ using Bessel functions \citep{Colwell92,colwell1993solving}:
\begin{equation}
\psi=M+\sum_{n=1}^\infty \dfrac{2J_n(ne)}{n}\sin (nM),
\end{equation}
where
\begin{equation}
J_n(x)=\sum_{k=0}^\infty \dfrac{(-1)^k x^{2k+n}}{2^{2k+n}k!(k+n)!}
\end{equation}
is the $n^{th}$--order \textit{Bessel function of the first kind}. For $e<1$, Bessel function asymptotics (\cite[See Sec. 8.4 of][]{watson1995treatise}) give
\begin{equation}
J_n(ne)\sim \dfrac{1}{\sqrt{2\pi\chi}}\dfrac{\exp (n\lambda)}{\sqrt{n}},\quad n\rightarrow\infty,
\end{equation}
where $\chi=\sqrt{1-e^2}$ is the aspect ratio of the ellipse and $\lambda$ is given by
\begin{equation}
\lambda=\chi+\dfrac{1}{2}\ln \dfrac{1-\chi}{1+\chi}.
\end{equation}
Since $\lambda$ is negative then $J_n(ne)\rightarrow 0$ as $n\rightarrow \infty$. Now we can write $\psi$ as
\begin{equation}
\psi=M+c_1e+c_2e^2+c_3e^3+c_4e^4+O(e^5),
\label{eq:anom1}
\end{equation}
where $c_1=\sin M,\;c_2=\dfrac{\sin(2M)}{2},\;c_3=\dfrac{3\sin(3M)-\sin M}{8},\;c_4=\dfrac{2\sin(4M)-\sin(2M)}{6}$.

Next we let the random variable $X\sim N(e,\sigma^2)$ in $(0,1)$ with mean eccentricity $e$ and variance $\sigma^2$ represent the fluctuating noise of eccentricity. Finally, we let $Y$ be the eccentric anomaly corresponding to $X$ through Kepler's equation, so that the approximate Bessel solution \eqref{eq:anom1} becomes
\begin{equation}
Y= M+c_1X+c_2X^2+c_3X^3+c_4X^4.
\label{eq:anom2}
\end{equation}
Table \ref{tab:mean_var_moments} gives the mean and variance of non-central moments.

\begin{table}[H]
\begin{center}
\caption{Mean and variance}\label{tab:mean_var_moments}
\begin{tabular}{ \thickvertical l|l|l \thickvertical} 
\thickhline
 & Mean & Variance\\
 \thickhline
 $X$ & $e$ & $\sigma^2$\\
 $X^2$ & $e^2+\sigma^2$ & $2\sigma^2(\sigma^2+2e^2)$\\
$X^3$ & $e^3+3e\sigma^2$ & $3\sigma^2(5\sigma^4+12\sigma^2e^2+3e^4)$\\
$X^4$ & $e^4+6e^2\sigma^2+3\sigma^4$ & $8\sigma^2(12\sigma^6+48\sigma^4e^2+21\sigma^2e^4+2e^6)$\\
\thickhline
\end{tabular}
\end{center}
\end{table}

\noindent The mean and variance of $Y$ are given by
\begin{equation}
\begin{array}{ll}
\varmathbb{E}(Y) & = M+c_1e+c_2(e^2+\sigma^2)+c_3(e^3+3e\sigma^2)+c_4(e^4+6e^2\sigma^2+3\sigma^4)\\
     & = (M+c_1X+c_2X^2+c_3X^3+c_4X^4 )+\sigma^2(c_2+3ec_3+6e^2c_4+3\sigma^2c_4)\\
     & =\psi+C_m\sigma^2, \qquad \text{and} \qquad\\
\text{Var}(Y)& = c_1^2\text{Var}(X)+c_2^2\text{Var}(X^2)+c_3^2\text{Var}(X^3)+c_4^2\text{Var}(X^4)\\
& = C_v \sigma^2.
\end{array}
\end{equation}
Therefore, $\varmathbb{E}(Y)\rightarrow \psi$ and $\text{Var}(Y)\rightarrow 0$ as $\sigma\rightarrow 0$. With sufficiently small values of $\sigma$, $Y$ is well approximated by a normal distribution $N(\psi,\sigma_Y^2)$. From the distribution of $Y$ to $\cos Y$, we use the Taylor series of the cosine function which is
\begin{equation}
\cos Y=\sum_{n=0}^\infty \dfrac{(-1)^n}{(2n)!}Y^{2n}=1-\dfrac{Y^2}{2!}+\dfrac{Y^4}{4!}-\dfrac{Y^6}{6!}+...
\end{equation}
When $\sigma_Y$ is sufficiently small, $\cos Y$ also has an approximately normal distribution with mean $\cos \psi$ and small variance. Using this fact, and the formula for the stochastic orbital distance $R=a(1-X\cos Y)$, we expect $R$ to have an approximately normal distribution with mean $r=a(1-e\cos \psi)$ and small variance $\sigma_R$ when $\sigma$ is small. With random variable $R^2$, we consider $R^2\sim\sigma_R^2\chi^2_{(1)}(\lambda_R)$, a non-central chi-squared distribution with degrees of freedom $k=1$ and noncentrality parameter $\lambda_R=\Big(\dfrac{r}{\sigma_R}\Big)^2$. According to Table \ref{tab:mean_var_moments}, $\varmathbb{E}(R^2)=r^2+\sigma_R^2$ and $\text{Var}(R^2)=2\sigma_R^2(\sigma_R^2+2r^2)$ converges to $r^2$ and 0 respectively when $\sigma_R\rightarrow 0$. Furthermore, the skewness of $R^2$ is $2\sqrt{2}\dfrac{1+3\lambda_R}{(1+2\lambda_R)^{\tfrac{3}{2}}}\rightarrow 0$ when $\lambda_R\rightarrow\infty$ or $\sigma_R\rightarrow 0$. Hence, when $\sigma$ is sufficiently small, we expect that $V=R^2$ has an approximately normal distribution with mean $r^2$ and small variance $\sigma_V$. 

\begin{figure}[H]
    \centering
    \includegraphics[width=\linewidth]{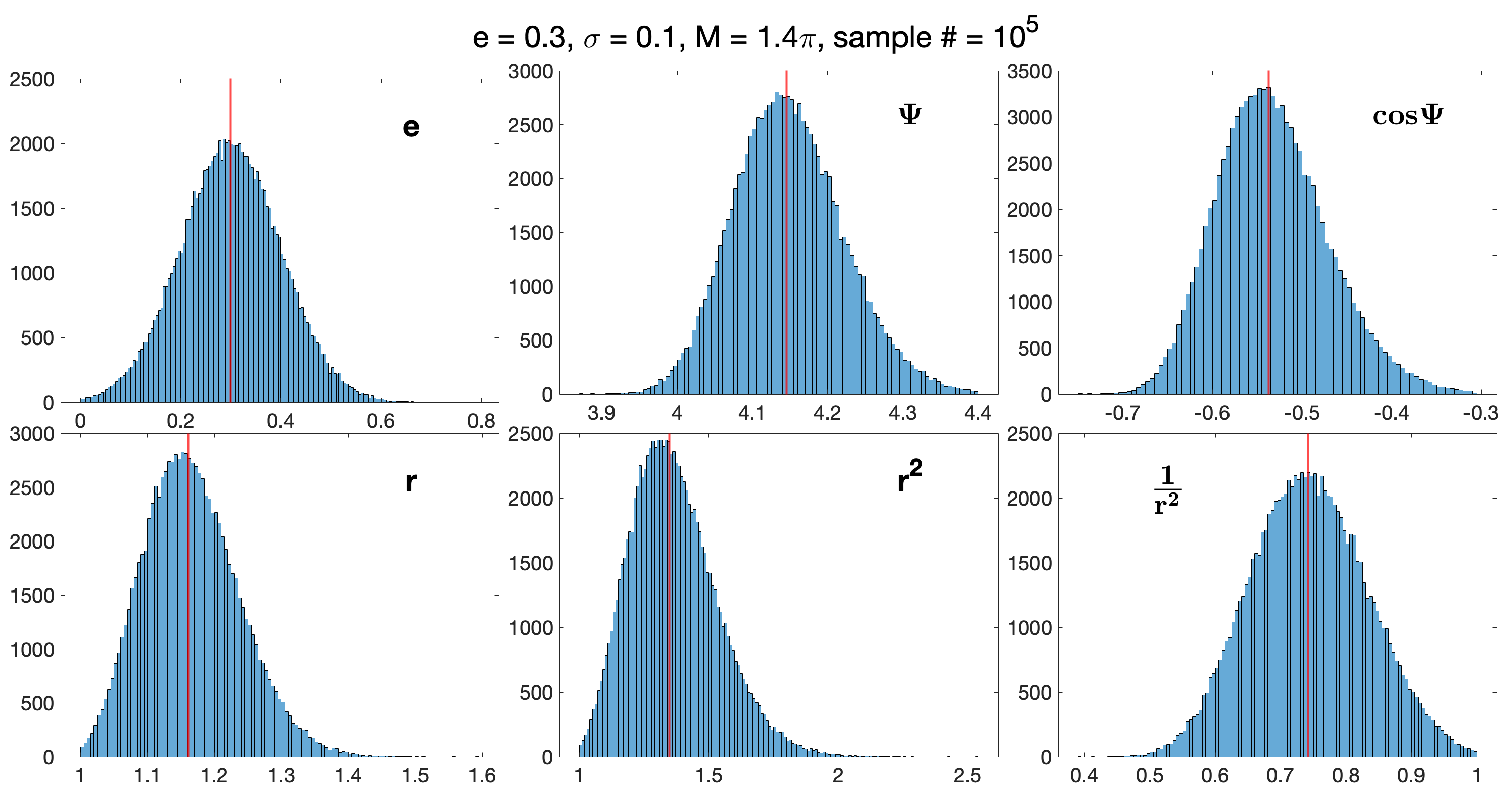}
    \caption{Distribution of noise values $e,\psi,\cos \psi,r,r^2$ and $\dfrac{1}{r^2}$ with $\sigma=0.1$ at $M=1.4\pi$. The red vertical lines are obtained by Kepler's equation and the formula for orbital distance.}
    \label{fig:noise_normal_2}
\end{figure}

\begin{figure}[H]
    \centering
    \includegraphics[width=\linewidth]{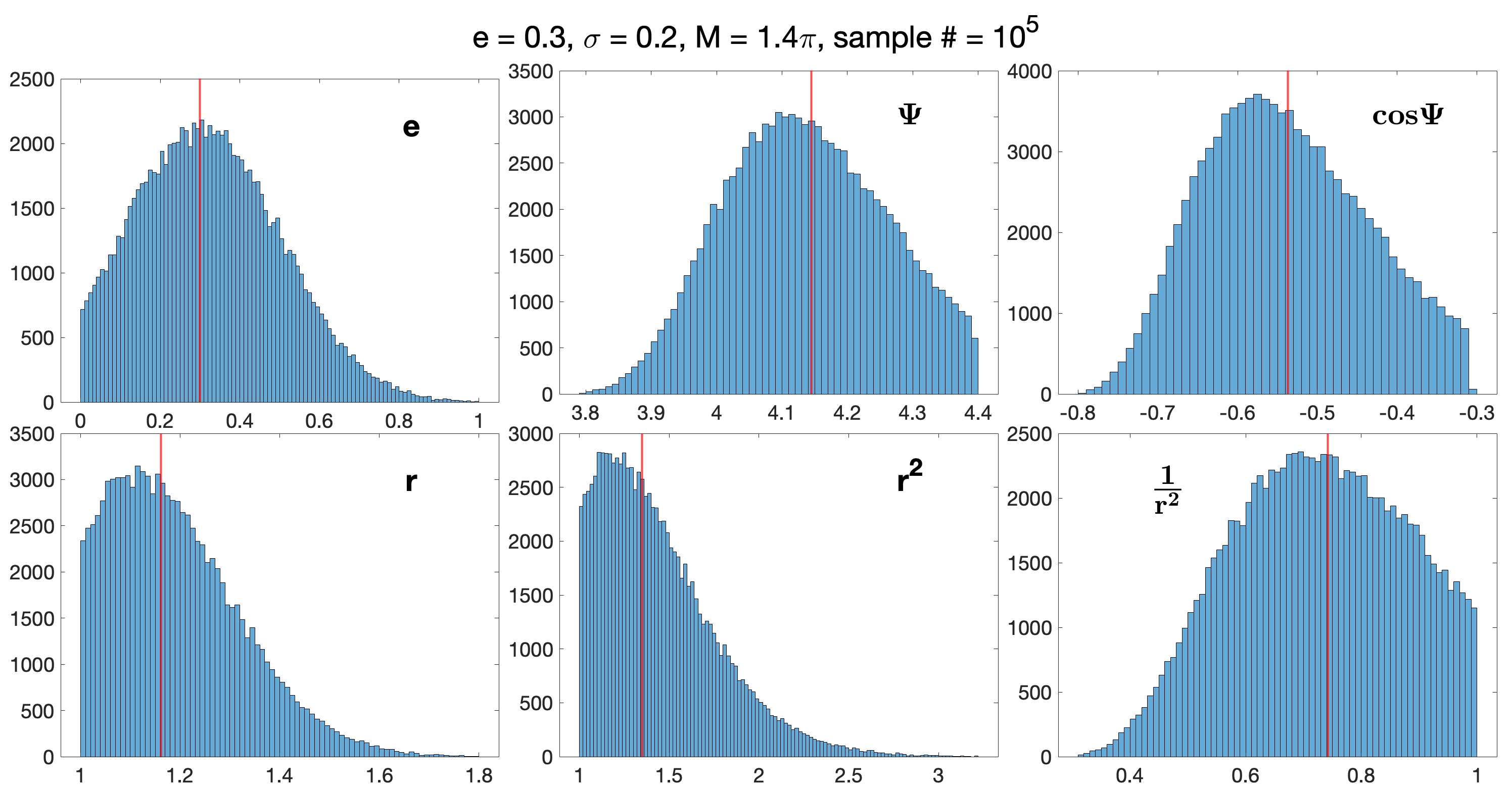}
    \caption{Distribution of noise values of $e,\psi,\cos \psi,r,r^2$ and $\dfrac{1}{r^2}$ with $\sigma=0.2$ at $M=1.4\pi$. The red vertical lines are obtained by Kepler's equation and the formula for orbital distance.}
    \label{fig:noise_sigma_larger_2}
\end{figure}

Finally, we consider the distribution of $U=\dfrac{1}{V}$, the random variable associated with the fluctuations of $\dfrac{1}{r^2}$ so that $U$ has a reciprocal normal distribution with probability density function
\[f_U(u)=\dfrac{1}{\sqrt{2\pi}\sigma_V u^2}\exp\Bigg[-\dfrac{1}{2}\bigg(\dfrac{1/u-r^2}{\sigma_V}\bigg)^2\Bigg]=\dfrac{1}{\sqrt{2\pi}\sigma_V u^2}\exp\Bigg[-\dfrac{1}{2}\bigg(\dfrac{u-\tfrac{1}{r^2}}{\sigma_V u\tfrac{1}{r^2}}\bigg)^2\Bigg].\]
We note that if $\sigma_V$ is sufficiently small, most of the values of $v$ are sharply centered about the mean $r^2>0$, and hence most of the values of $u=\dfrac{1}{v^2}$ are sharply centered about $\dfrac{1}{r^2}>0$. Given this, we can assign standard deviation $\sigma_U=\sigma_V\dfrac{1}{r^4}$, such that $\sigma_Vu^2\approx\sigma_Vu\dfrac{1}{r^2}\approx\sigma_U$. Therefore, the probability density function
\[f_U(u)\approx\dfrac{1}{\sqrt{2\pi}\sigma_U}\exp\Bigg[-\dfrac{1}{2}\bigg(\dfrac{u-\tfrac{1}{r^2}}{\sigma_U}\bigg)^2\Bigg],\]
approximates a normal distribution with mean $\dfrac{1}{r^2}$ and variance $\sigma_U \ll1$. We illustrate this case in Figure \ref{fig:noise_normal_2} when $e=0.3$ and $\sigma=0.1$ at $M=1.4\pi$, and the sample size is 100,000. The histograms of the noise values have 
symmetric normal distributions around the mean values (red vertical lines) obtained from Kepler's equations and formulas of elliptic orbits, with slight skewness for $r$ and $r^2$. On the other hand, when $\sigma$ is larger, we do not obtain Gaussian distributions, as shown in Figure \ref{fig:noise_sigma_larger_2} for $\sigma=0.2$, where the uncertainty in noise values is larger.




\section{Supplemental Material\label{sec:SM}}

Here we provide a quantitative argument for the high-frequency fluctuations in eccentricity based on circumbinary planets Kepler-16b, Kepler-34b and Kepler-35b.   We then collect additional figures to supplement the main text.

\section*{Eccentricity Fluctuations of Circumbinary Planets}

Using the values in Table~2 of \citet{LeungLee2013AnalyticTheory}, particularly the {\em apsidal}, {\em guiding-center} and {\em epicyclic or free eccentricity} frequencies, we discuss the origin of high frequency, or short period, osculating eccentricity as seen in their Figures 1, 4, 5 and 7.  

\citet{LeungLee2013AnalyticTheory} represent the planetary motion by superposing the circular motion of a guiding center, the forced oscillations, originating in the non-axisymmetric components of the binary’s potential, the epicyclic and vertical motions.



We determine the frequencies $n_0$ and $\kappa_0$ (apsidal), $n_0$ and $\nu_0$ (nodal), using the values in their Table~2 of the Keplerian mean frequency $n_K$ = yr$^{-1}$, and the ratios $n_0/n_K$, $\kappa_0/n_K$ and $\nu_0/n_K$ viz. 
\begin{equation}
n_0 = \left(\frac{n_0}{n_K}\right)n_K\qquad \textrm{and} \qquad
\kappa_0 = \left(\frac{\kappa_0}{n_K}\right)n_K \qquad \textrm{and} \qquad \nu_0 = \left(\frac{\nu_0}{n_K}\right)n_K,
\nonumber
\end{equation}
to calculate the guiding center, epicyclic or free eccentricity, and apsidal periods, $P_0$, $P_\kappa$ and $P_\text{aps}$ respectively, for Kepler-16 b, 34 b, 35 b as follows.\\

{\bf Relevant periods for Kepler-16 b}.\\

(1) Guiding-center period:
\begin{equation}
P_0 = \frac{2\pi}{n_0} \approx \frac{2\pi}{10.1531} \approx 0.619~{\rm yr} \approx 226~{\rm days}.
\end{equation}

(2) Epicyclic/free-eccentricity period:
\begin{equation}
P_\kappa = \frac{2\pi}{\kappa_0} \approx \frac{2\pi}{10.0041} \approx 0.628~{\rm yr} \approx 229~{\rm days}
\end{equation}

(3) Apsidal (beat) period:
\begin{equation}
P_{\rm aps} = \frac{2\pi}{|n_0-\kappa_0|} \approx \frac{2\pi}{0.1490} \approx 42.2~{\rm yr}.
\end{equation}

{\bf Relevant periods for Kepler-34 b}.\\

Using
\begin{equation}
n_0 = 8.0512\times 1.00423 \approx 8.0853~{\rm yr^{-1}},
\qquad
\kappa_0 = 8.0512\times 0.99567 \approx 8.0163~{\rm yr^{-1}}.
\nonumber
\end{equation}

(1) Guiding-center period:
\begin{equation}
P_0 \approx \frac{2\pi}{8.0853} \approx 0.777~{\rm yr} \approx 284~{\rm days},
\end{equation}

(2) Epicyclic/free-eccentricity period:
\begin{equation}
P_\kappa \approx \frac{2\pi}{8.0163} \approx 0.784~{\rm yr} \approx 286~{\rm days},
\end{equation}

(3) Apsidal (beat) period:
\begin{equation}
P_{\rm aps} = \frac{2\pi}{|n_0-\kappa_0|},\quad
|n_0-\kappa_0|\approx 0.0689~{\rm yr^{-1}},
\quad
P_{\rm aps}\approx \frac{2\pi}{0.0689} \approx 91.1~{\rm yr}.
\end{equation}

{\bf Relevant periods for Kepler-35 b}.\\

Using 
\begin{equation}
n_0 = 17.8875\times 1.00838 \approx 18.0374~{\rm yr^{-1}},
\qquad
\kappa_0 = 17.8875\times 0.99119 \approx 17.7299~{\rm yr^{-1}}.
\nonumber
\end{equation}

(1) Guiding-center period:
\begin{equation}
P_0 \approx \frac{2\pi}{18.0374} \approx 0.348~{\rm yr} \approx 127~{\rm days},
\end{equation}

(2) Epicyclic/free-eccentricity period:
\begin{equation}
P_\kappa \approx \frac{2\pi}{17.7299} \approx 0.354~{\rm yr} \approx 129~{\rm days},
\end{equation}

(3) Apsidal (beat) period:
\begin{equation}
P_{\rm aps} = \frac{2\pi}{|n_0-\kappa_0|},\quad
\quad
|n_0-\kappa_0|\approx 0.3075~{\rm yr^{-1}},
\quad
P_{\rm aps}\approx \frac{2\pi}{0.3075} \approx 20.4~{\rm yr}.
\end{equation}
 
The main point is that the slower apsidal beats
\begin{equation}
P_{\rm aps}\approx 42~{\rm yr},\;91~{\rm yr},\;20~{\rm yr}, 
\end{equation}
control the long term modulation/envelope upon which high frequency fluctuations occur.  
These dominant fast periods in the solution of \citet{LeungLee2013AnalyticTheory} are $P_0$ and $P_\kappa$ . From the above, these are sub-year across Kepler-16 b, 34 b, 35 b:
\begin{equation}
P_0,\,P_\kappa \approx (0.35\text{ to }0.78)~{\rm yr}, 
\end{equation}
so when eccentricity is computed from the instantaneous state (osculating elements), it is natural to observe high frequency eccentricity oscillations on roughly subyear timescales because of beating between $n_0$ and $\kappa_0$, plus any additional forced harmonics. 

Additional effects in these and other circumbinary systems (Kepler-38, Kepler-64, Kepler-413) leading to multiple timescale fluctuations include, inter-alia, $p:1$ mean-motion resonances, with the possibility of bounded oscillations, or libration of resonance angles, eccentricity pumping, and chaotic overlap near the inner stability limit, which are more likely for planets sitting close to, or migrating through, these commensurabilities, especially for higher binary eccentricity.

\color{black}

\section*{Supplementary Figures}

\begin{figure}[H]
    \centering
    \begin{subfigure}{0.49\textwidth}
        \centering
        \includegraphics[width=\textwidth]{Figures/final_con_path1k_e01.png}
    \end{subfigure}
    \begin{subfigure}{0.49\textwidth}
        \centering
        \includegraphics[width=\textwidth]{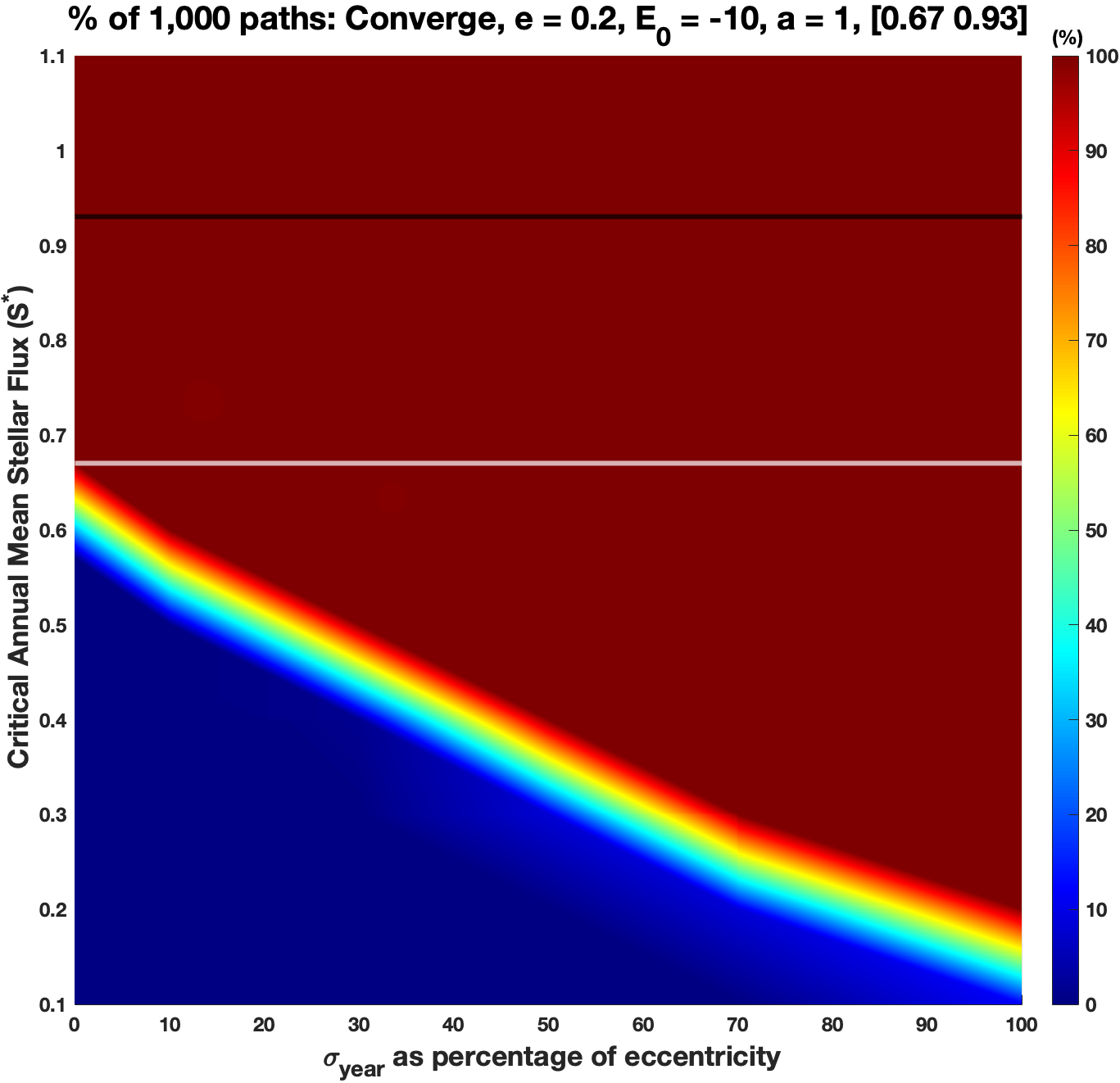}
    \end{subfigure}

    \begin{subfigure}{0.49\textwidth}
        \centering
        \includegraphics[width=\textwidth]{Figures/final_con_path1k_e03.png}
    \end{subfigure}
    \begin{subfigure}{0.49\textwidth}
        \centering
        \includegraphics[width=\textwidth]{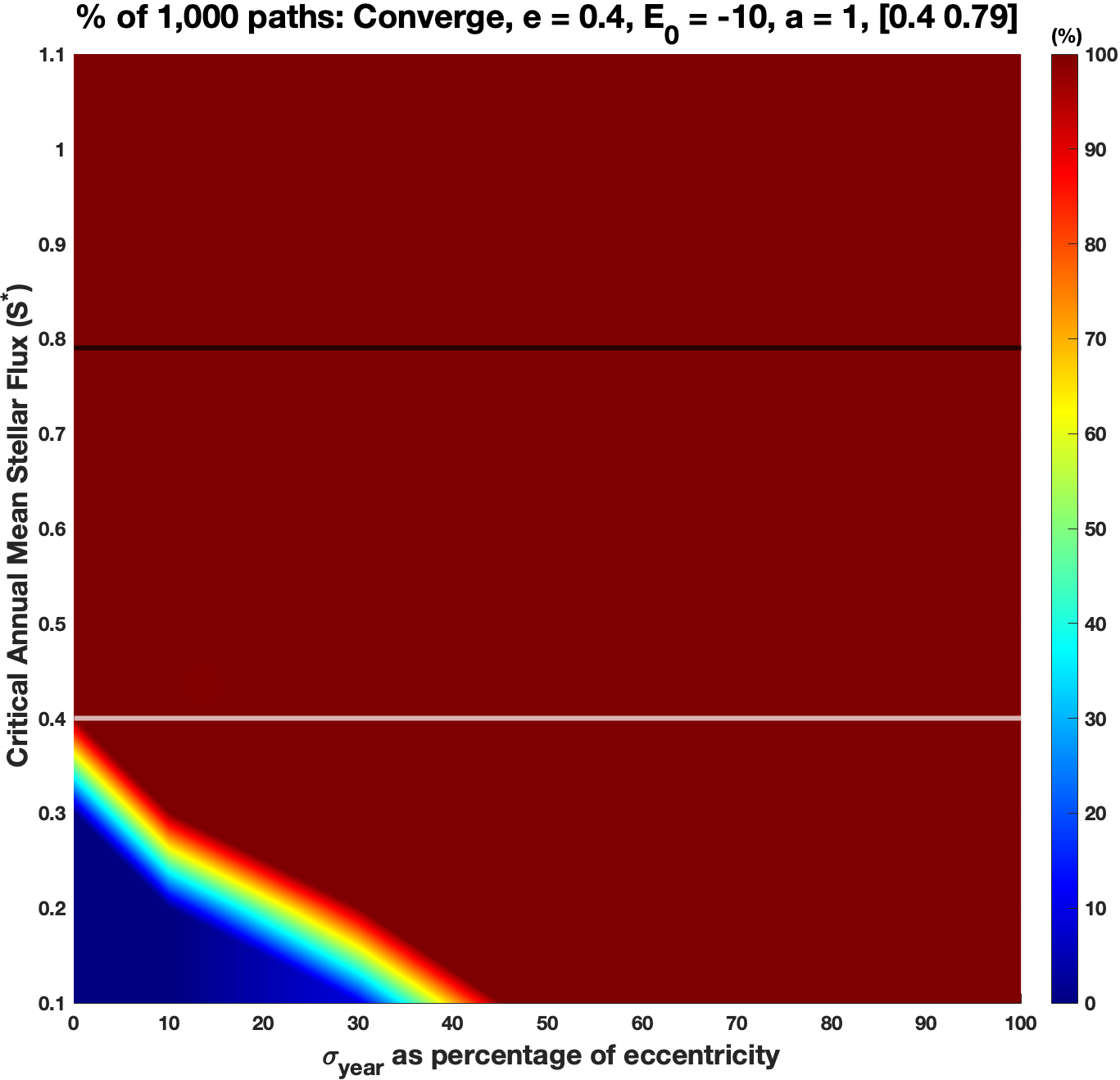}
    \end{subfigure}

    \begin{subfigure}{0.49\textwidth}
        \centering
        \includegraphics[width=\textwidth]{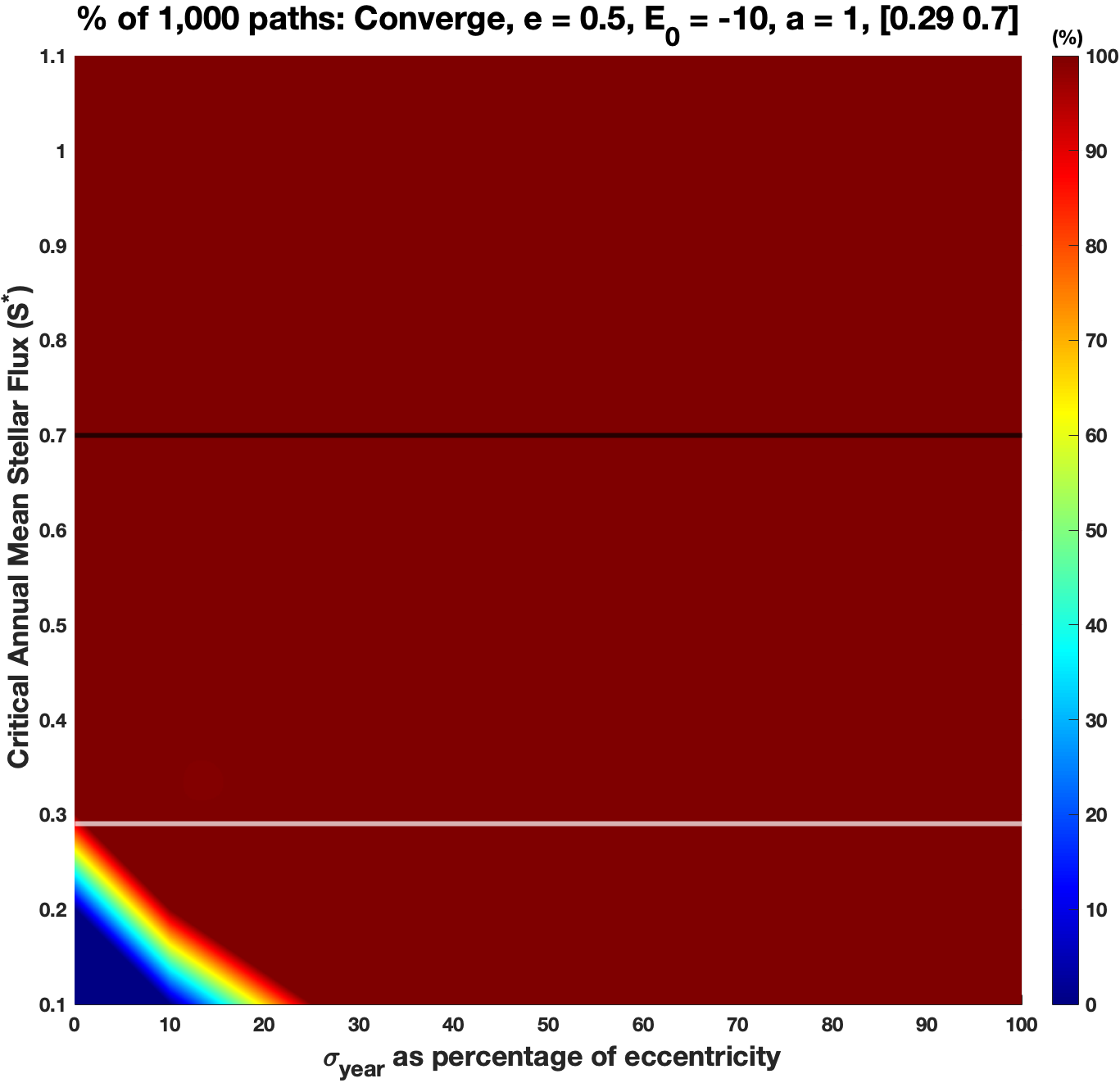}
    \end{subfigure}
    \begin{subfigure}{0.49\textwidth}
        \centering
        \includegraphics[width=\textwidth]{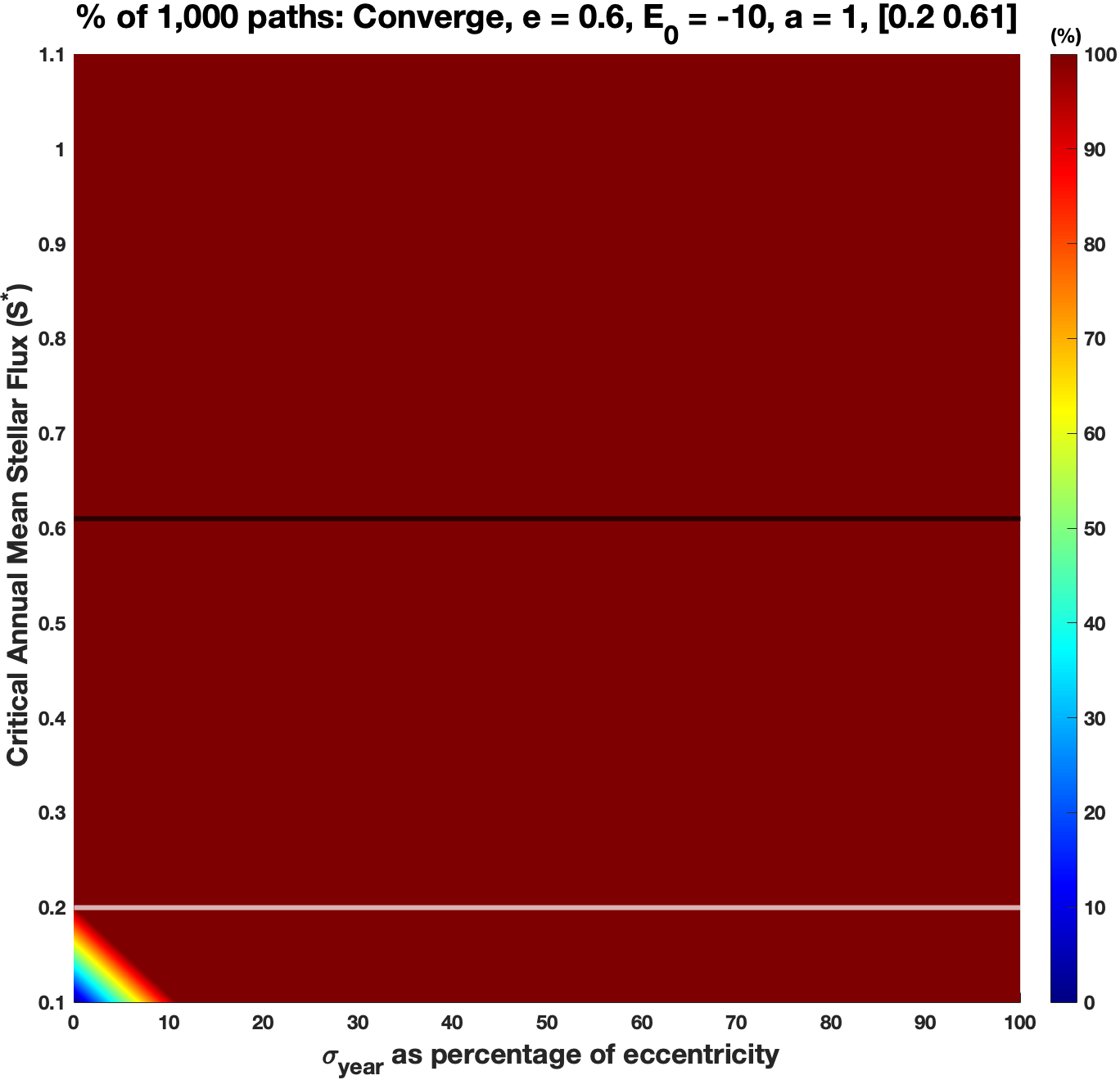}
    \end{subfigure}
    \caption{Percentage of 1,000 stochastic paths that converge when $e=0.1,\,0.2,\,0.3,\,0.4,\,0.5,\,0.6$ shown as a color map. 
    The two bracketed values in the title of each sub-figure are the critical deterministic values of $S^*$ for convergence and bifurcation given in Table A.2 of the main paper.
    The white and black horizontal lines are the convergence and bifurcation values respectively.}
    \label{fig:percent_con_1000paths_SM}
\end{figure}

\begin{figure}[H]
    \centering
    \begin{subfigure}{0.49\textwidth}
        \centering
        \includegraphics[width=\textwidth]{Figures/final_yr_path1k_e01.png}
    \end{subfigure}
    \begin{subfigure}{0.49\textwidth}
        \centering
        \includegraphics[width=\textwidth]{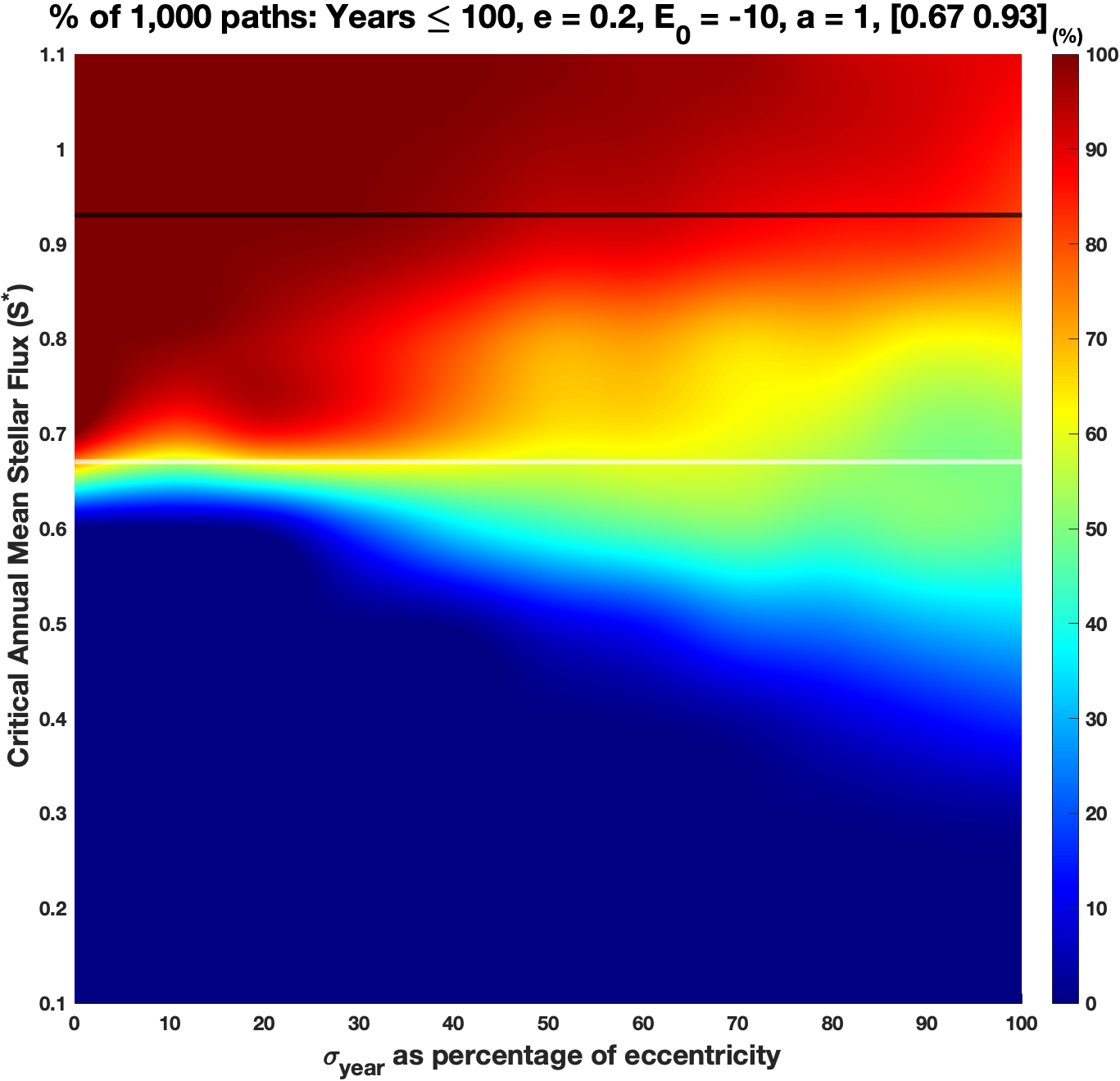}
    \end{subfigure}

    \begin{subfigure}{0.49\textwidth}
        \centering
        \includegraphics[width=\textwidth]{Figures/final_yr_path1k_e03.png}
    \end{subfigure}
    \begin{subfigure}{0.49\textwidth}
        \centering
        \includegraphics[width=\textwidth]{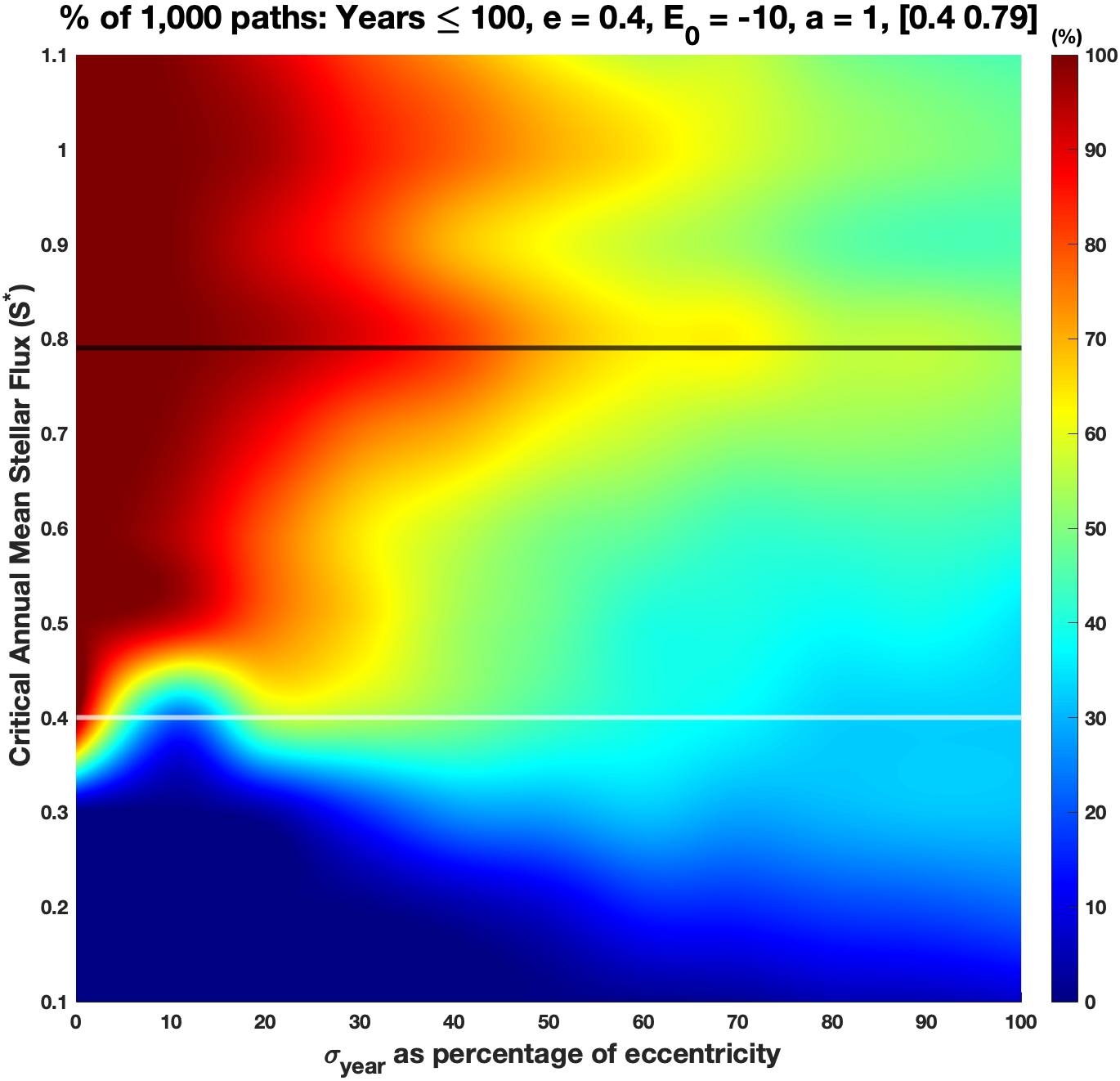}
    \end{subfigure}

    \begin{subfigure}{0.49\textwidth}
        \centering
        \includegraphics[width=\textwidth]{Figures/final_yr_path1k_e05.png}
    \end{subfigure}
    \begin{subfigure}{0.49\textwidth}
        \centering
        \includegraphics[width=\textwidth]{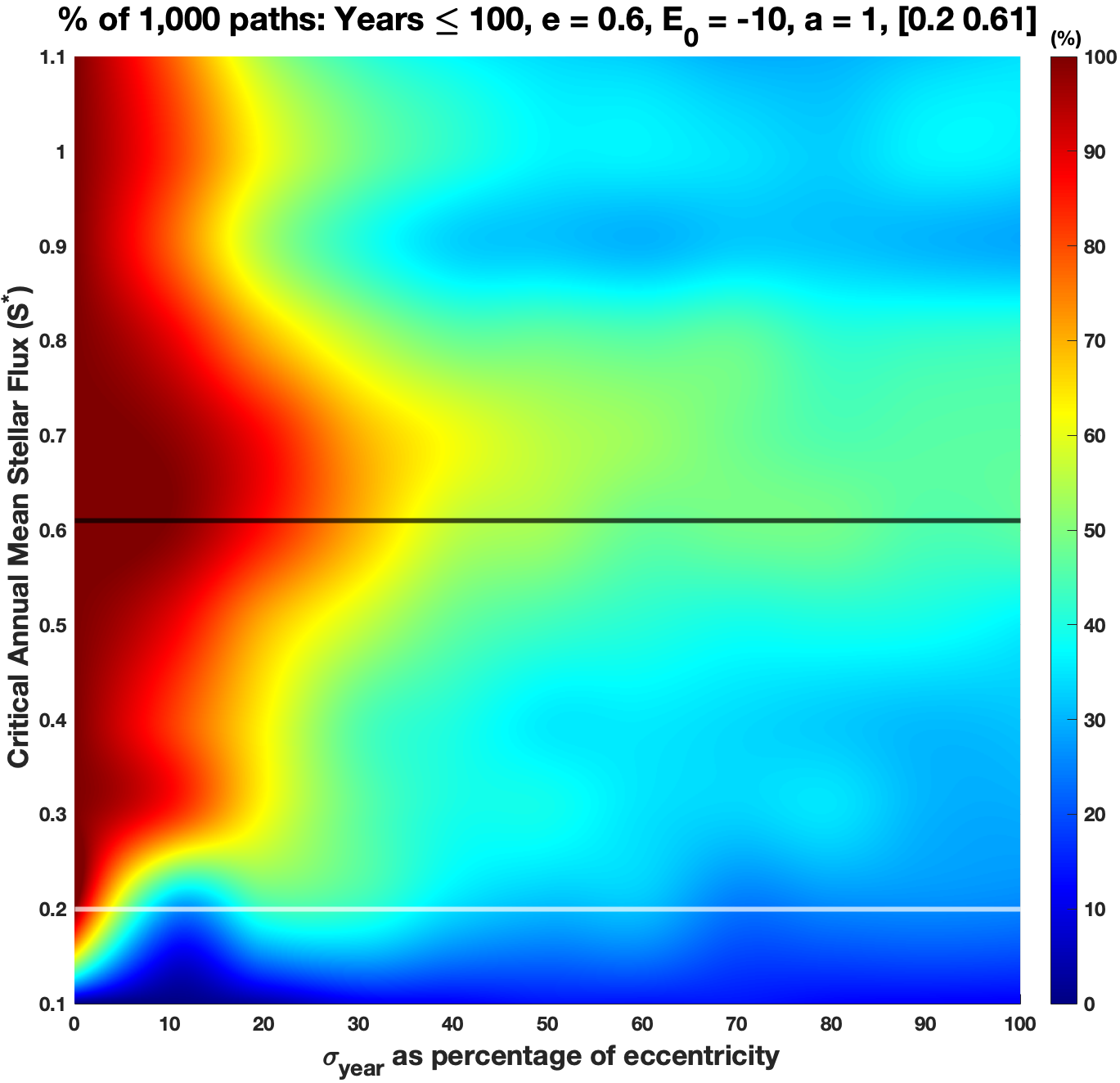}
    \end{subfigure}
    \caption{Percentage of 1,000 stochastic paths that converge within 100 years when $e=0.1,\,0.2,\,0.3,\,0.4,\,0.5,\,0.6$ shown as a color map. 
    The two bracketed values in the title of each sub-figure are the critical deterministic values of $S^*$ for convergence and bifurcation given in Table A.2 of the main paper.
    The white and black horizontal lines are the convergence and bifurcation values respectively.}
    \label{fig:percent_yr_1000paths_SM}
\end{figure}

\begin{figure}[H]
    \centering
    \begin{subfigure}{0.49\textwidth}
        \centering
        \includegraphics[width=\textwidth]{Figures/final_bif_path1k_e01.png}
    \end{subfigure}
    \begin{subfigure}{0.49\textwidth}
        \centering
        \includegraphics[width=\textwidth]{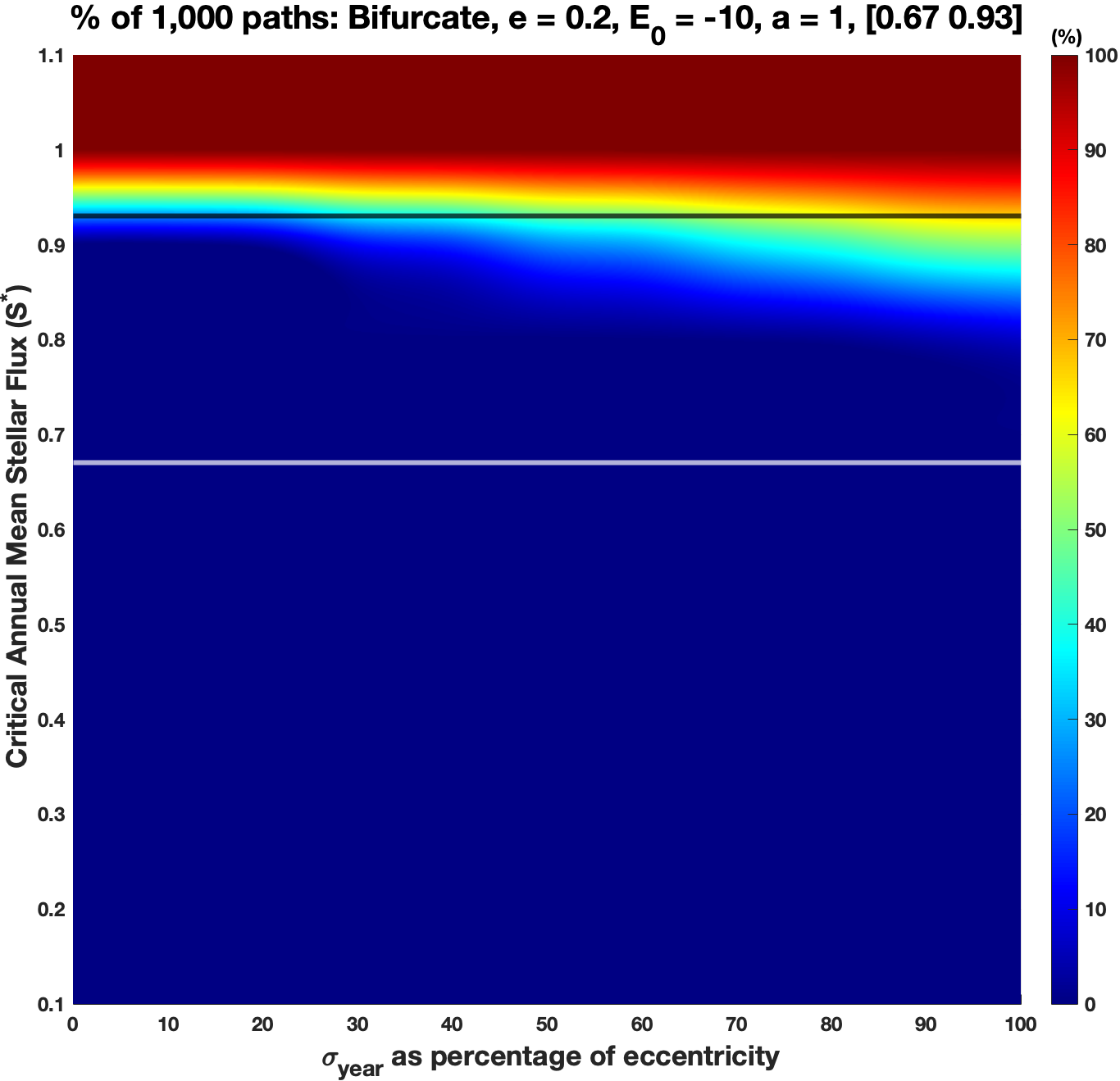}
    \end{subfigure}

    \begin{subfigure}{0.49\textwidth}
        \centering
        \includegraphics[width=\textwidth]{Figures/final_bif_path1k_e03.png}
    \end{subfigure}
    \begin{subfigure}{0.49\textwidth}
        \centering
        \includegraphics[width=\textwidth]{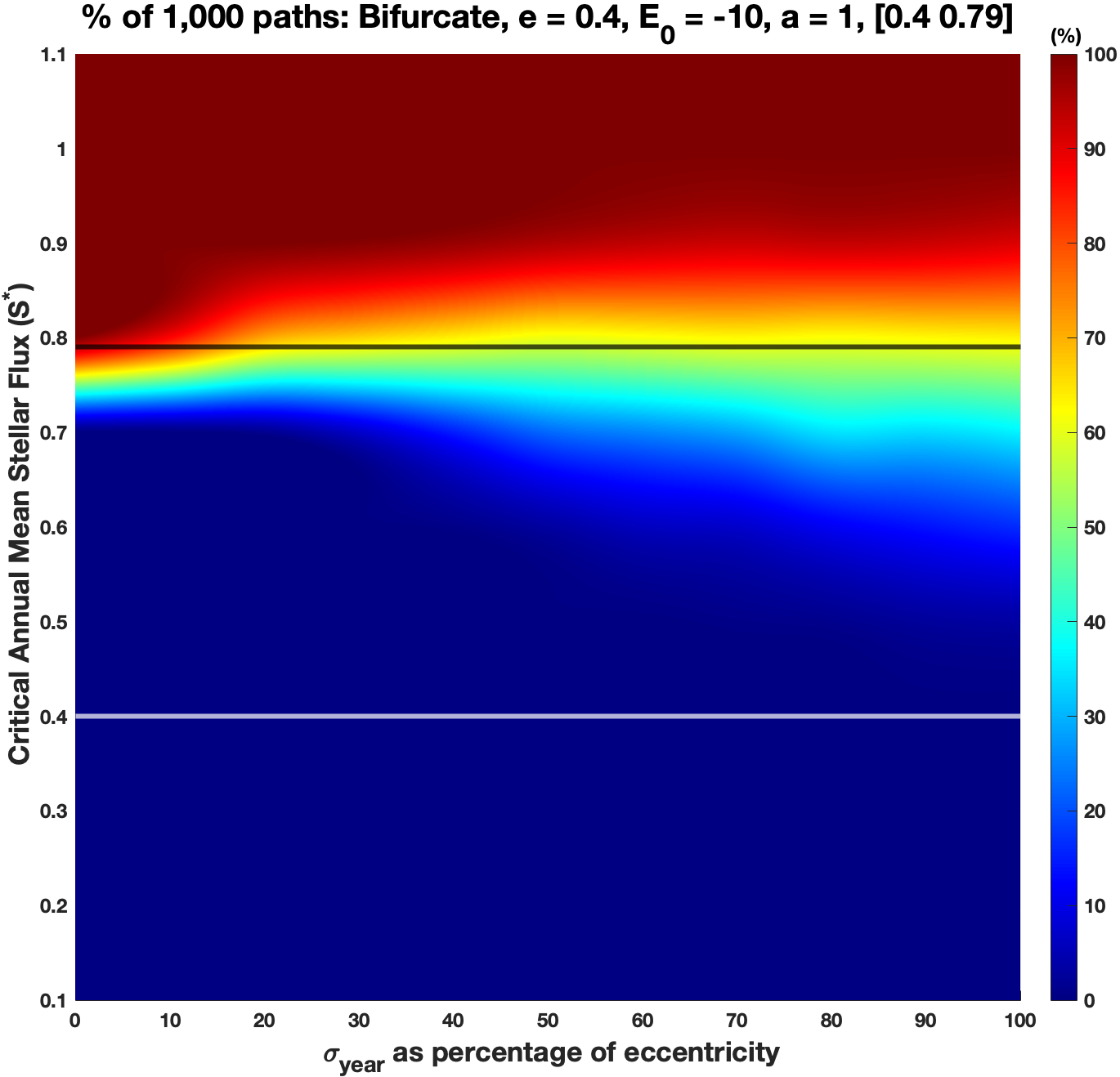}
    \end{subfigure}

    \begin{subfigure}{0.49\textwidth}
        \centering
        \includegraphics[width=\textwidth]{Figures/final_bif_path1k_e05.png}
    \end{subfigure}
    \begin{subfigure}{0.49\textwidth}
        \centering
        \includegraphics[width=\textwidth]{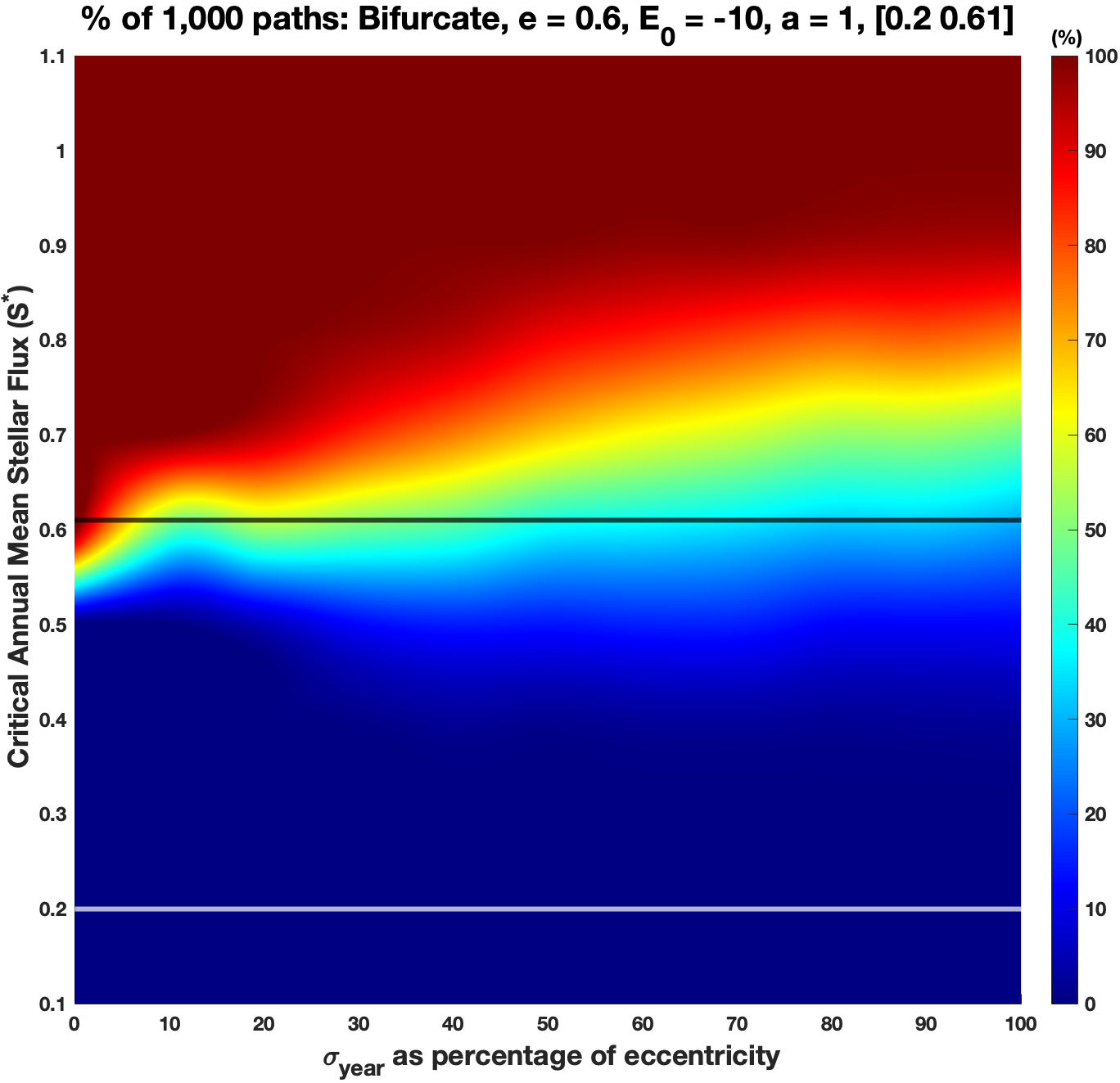}
    \end{subfigure}
    \caption{The percentage of bifurcations in 1,000 stochastic paths when $e=0.1,\,0.2,\,0.3,\,0.4,\,0.5,\,0.6$ shown as a color map. The two bracketed values in the title of each sub-figure are the critical deterministic values of $S^*$ for convergence and bifurcation given in Table A.2 of the main paper.
    The white and black horizontal lines are the convergence and bifurcation values respectively.}
    \label{fig:percent_bif_1000paths_SM}
\end{figure}

\section*{Eccentricity of exoplanets in the NASA database}\label{se:NASA_database}

\begin{figure}[H]
    \centering
    \begin{subfigure}{0.49\textwidth}
        \centering
        \includegraphics[width=\textwidth]{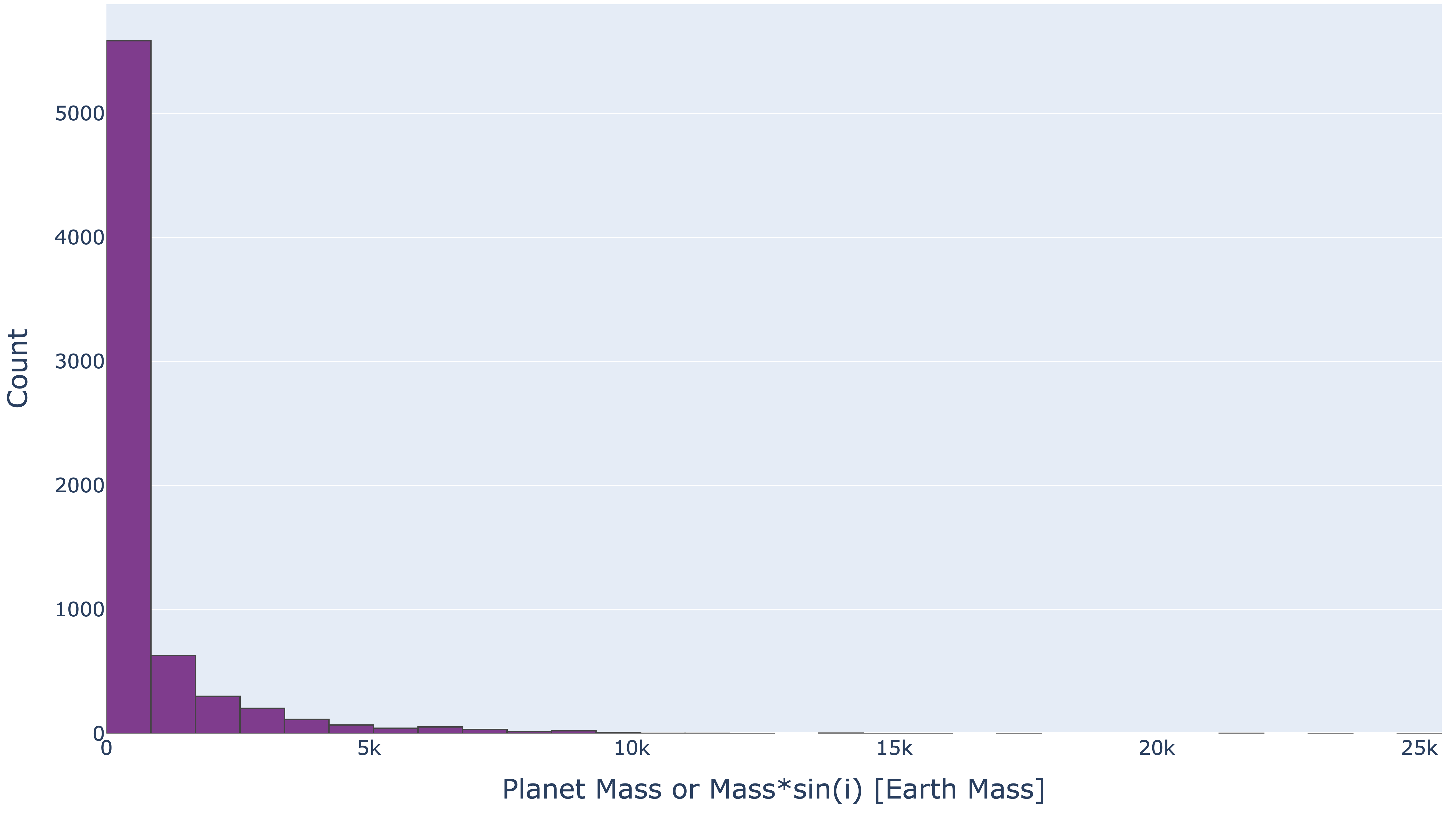}
    \end{subfigure}
    \begin{subfigure}{0.49\textwidth}
        \centering
        \includegraphics[width=\textwidth]{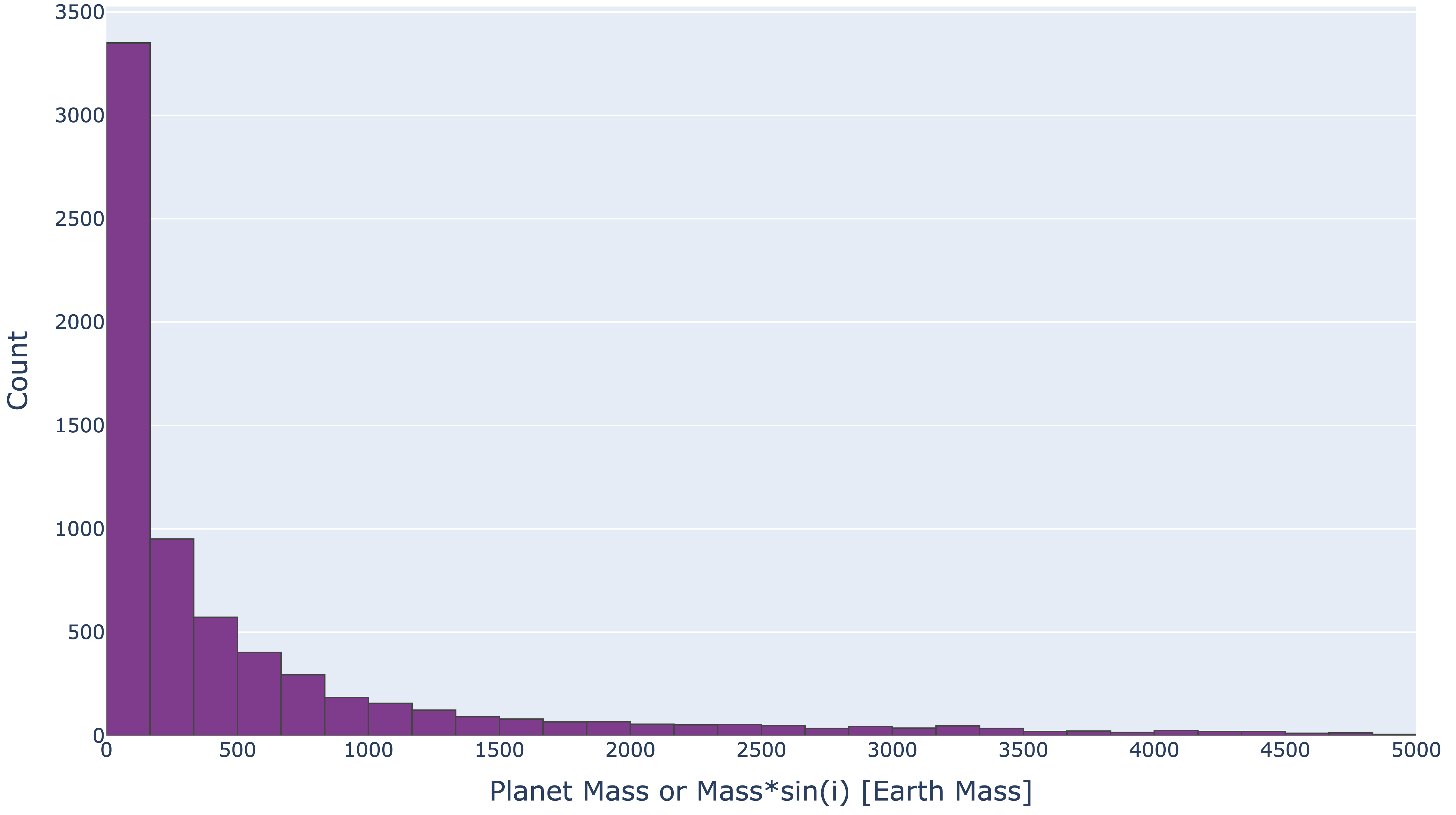}
    \end{subfigure}
    \caption{Masses of exoplanets in the \cite{nasa} database. The left figure is the histogram of all confirmed exoplanets, and the right figure is the histogram of all exoplanets whose mass is capped at 5000 Earth Masses.}
\end{figure}

\begin{figure}[H]
    \centering
    \begin{subfigure}{0.49\textwidth}
        \centering
        \includegraphics[width=\textwidth]{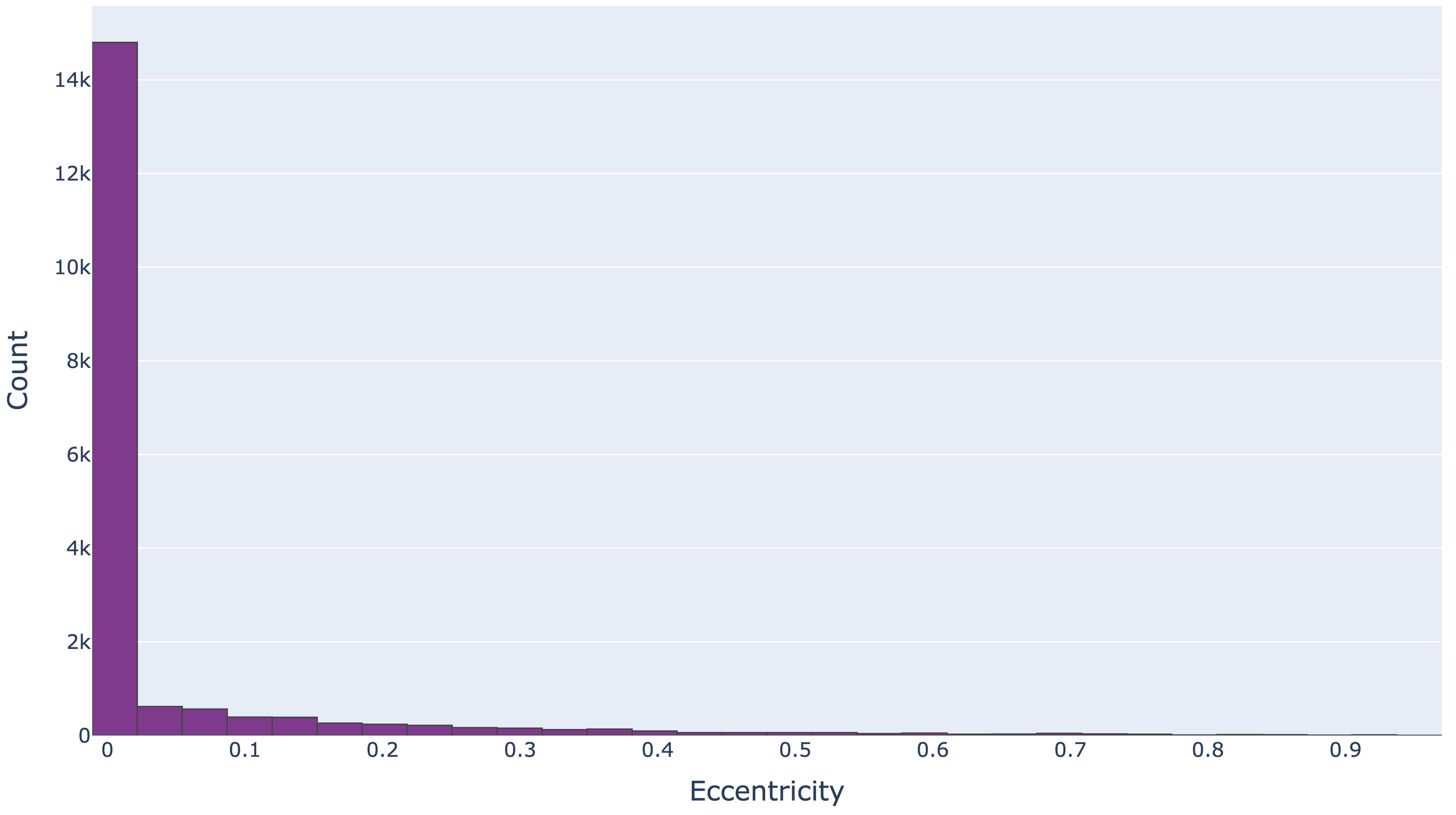}
    \end{subfigure}
    \begin{subfigure}{0.49\textwidth}
        \centering
        \includegraphics[width=\textwidth]{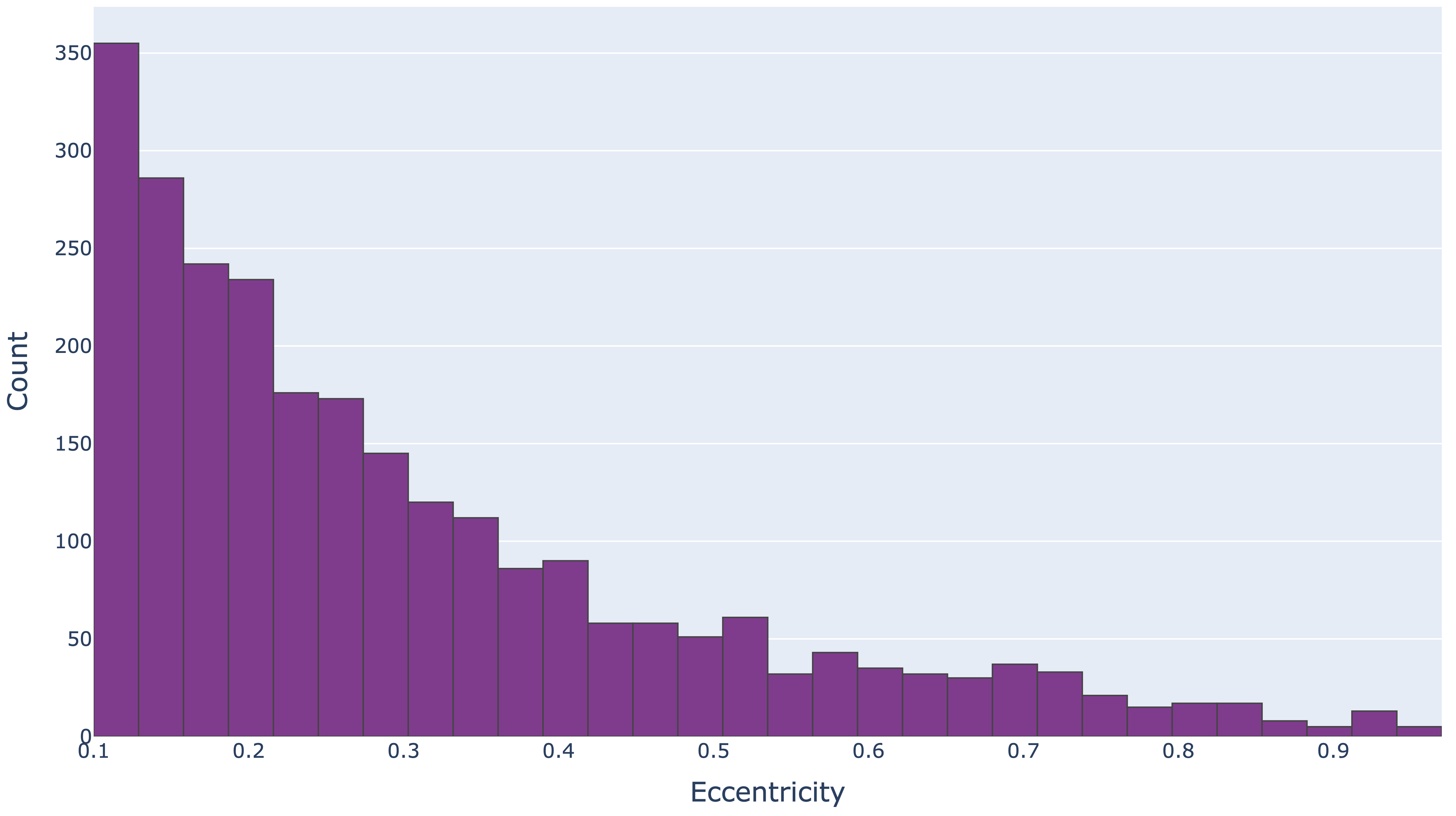}
    \end{subfigure}
    \caption{Eccentricities of exoplanets in the \cite{nasa} database. The left figure is the histogram of all confirmed exoplanets, and the right figure is the histogram of all exoplanets whose eccentricity is larger than 0.1. For reference, Earth's value is 0.0167.}
\end{figure}

\begin{figure}[H]
    \centering
    \begin{subfigure}{\textwidth}
        \centering
        \includegraphics[width=\textwidth]{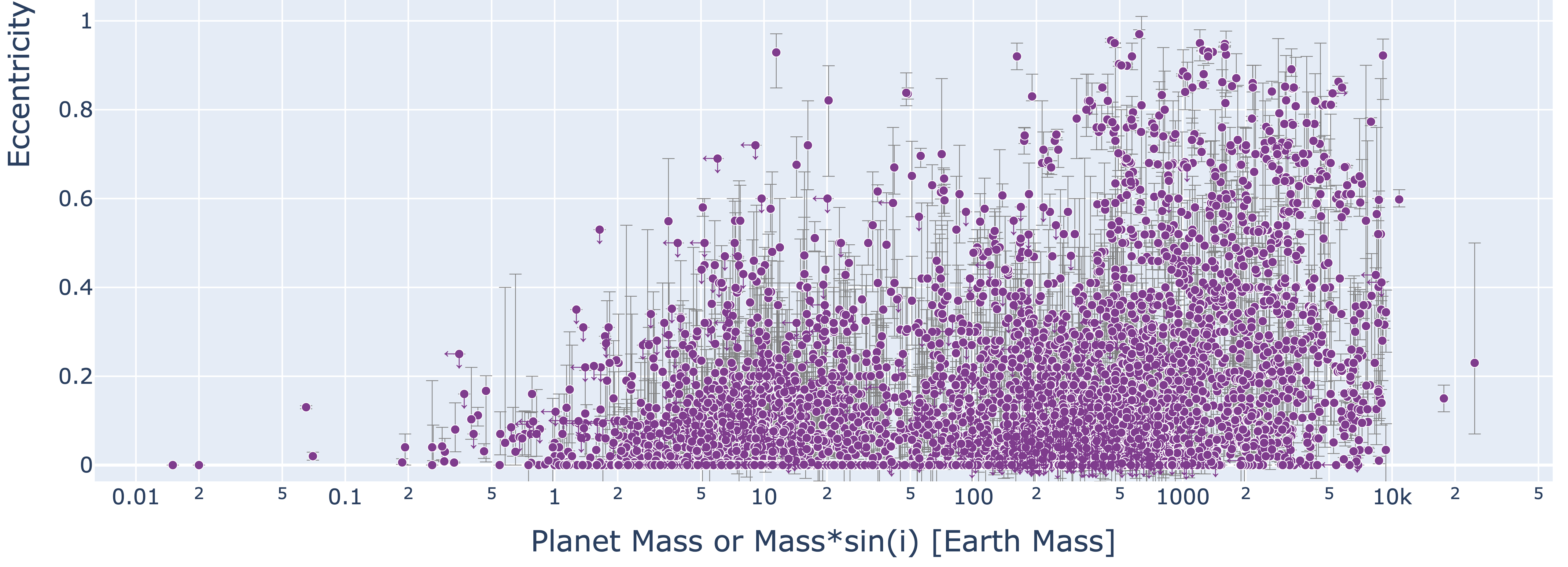}
    \end{subfigure}
    
    \begin{subfigure}{\textwidth}
        \centering
        \includegraphics[width=\textwidth]{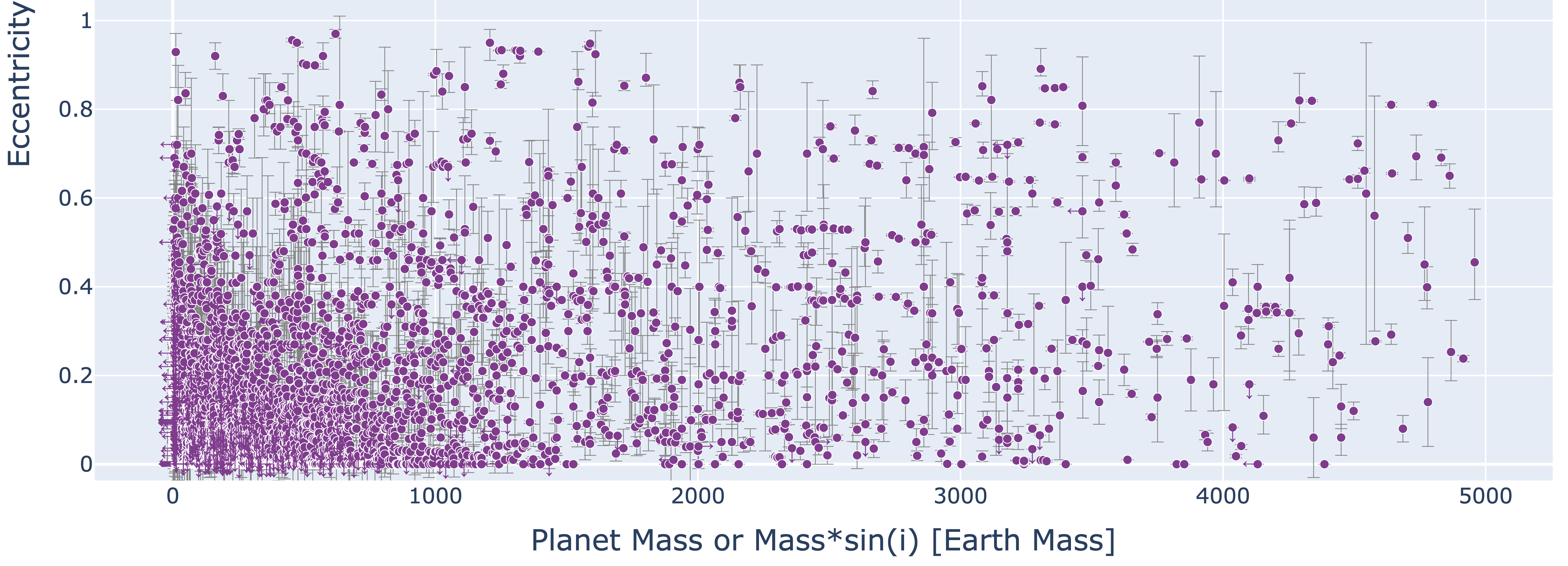}
    \end{subfigure}
    \caption{Eccentricity (with error bars for uncertain observations) vs Earth Mass of exoplanets in the \cite{nasa} database. The top figure is for all exoplanets, and the bottom is for exoplanets whose mass is capped at 5000 Earth Masses. The horizontal axis in the top figure is scaled in powers of 10, whereas the bottom figure is plotted in a linear scale.}
\end{figure}

\end{document}